\documentclass[sigconf]{acmart}

\AtBeginDocument{%
  }

\acmYear{2024}\copyrightyear{2024}
\acmConference[ASIA CCS '24]{ACM Asia Conference on Computer and Communications Security}{July 1--5, 2024}{Singapore, Singapore}
\acmBooktitle{ACM Asia Conference on Computer and Communications Security (ASIA CCS '24), July 1--5, 2024, Singapore, Singapore}
\acmDOI{10.1145/3634737.3656287}
\acmISBN{979-8-4007-0482-6/24/07}

\usepackage{color}

\newcommand\Tstrut{\rule{0pt}{2.4ex}}
\usepackage{wasysym}
\usepackage{threeparttable}
\usepackage{multirow}
\newcommand{\tabgod}{\CIRCLE}
\newcommand{\tabmed}{\LEFTcircle}
\newcommand{\tabbad}{\Circle}
\usepackage{tikz}
\usetikzlibrary{shapes.geometric}
\usetikzlibrary{positioning}
\usetikzlibrary{calc}
\tikzstyle{nn_arrow} = [black, ->, thick]
\DeclareMathOperator*{\argmax}{arg\,max}
\DeclareMathOperator*{\argmin}{arg\,min}
\usepackage{colortbl}

\newcommand\copyrightnotice[1]{
	\begin{tikzpicture}[remember picture,overlay]
		\node[anchor=south,yshift=20pt] at (current page.south) {\fbox{\parbox{\dimexpr\textwidth-\fboxsep-\fboxrule\relax}{#1}}};
	\end{tikzpicture}
}

\begin{document}

\title[Towards Robust Domain Generation Algorithm Classification]{Towards Robust Domain Generation Algorithm Classification}



\author{Arthur Drichel}
\email{drichel@itsec.rwth-aachen.de}
\affiliation{%
			\institution{RWTH Aachen University}
			\city{}
			\country{}
		}
	
\author{Marc Meyer}
\email{marc.meyer@rwth-aachen.de}
\affiliation{%
		\institution{RWTH Aachen University}
		\city{}
		\country{}
	}
	
\author{Ulrike Meyer}
\email{meyer@itsec.rwth-aachen.de}
\affiliation{%
			\institution{RWTH Aachen University}
			\city{}
			\country{}
	}

\renewcommand{\shortauthors}{Drichel et al.}

\begin{abstract}
In this work, we conduct a comprehensive study on the robustness of domain generation algorithm (DGA) classifiers.
We implement 32 white-box attacks, 19 of which are very effective and induce a false-negative rate (FNR) of $\approx$ 100\% on unhardened classifiers.
To defend the classifiers, we evaluate different hardening approaches and propose a novel training scheme that leverages adversarial latent space vectors and discretized adversarial domains to significantly improve robustness.
In our study, we highlight a pitfall to avoid when hardening classifiers and uncover training biases that can be easily exploited by attackers to bypass detection, but which can be mitigated by adversarial training (AT).
In our study, we do not observe any trade-off between robustness and performance, on the contrary, hardening improves a classifier's detection performance for known and unknown DGAs.
We implement all attacks and defenses discussed in this paper as a standalone library, which we make publicly available\footnote{\label{sc}\url{https://gitlab.com/rwth-itsec/robust-dga-detection}} to facilitate hardening of DGA classifiers.
\end{abstract}


%

\maketitle

\section{Introduction}
\label{sec:introduction}
Botnets are collections of malware-infected hosts (bots) that an adversary (botnet master) can control remotely to perform malicious deeds such as distributed denial-of-service (DDoS) attacks, sending spam, or stealing personal data.
Thereby, botnets pose a serious threat to Internet users and organizations. 
For instance, the Mantis botnet~\cite{Yoachimik2022} was able to launch a DDoS attack peaking at 26 Million HTTPS requests/s using only 5000 bots.
And as of late August 2023, Cloudflare observed record-breaking DDoS attacks surpassing 201 million requests/s with only 20 thousand bots (a relatively modest size by the standards of modern botnets)~\cite{Pardue2023}.
Other botnets have been shown to span more than one million bots~\cite{Plohmann2016,Rossow2013}. 

The bots must regularly connect to a command-and-control (C2) server operated by the botnet master in order to receive commands or exfiltrate sensitive data.
To this end, modern botnets rely on domain generation algorithms (DGAs), which are pseudo-random algorithms that generate a huge amount of algorithmically generated domains (AGDs) that serve as rendezvous points between bots and their masters.
The botnet master is aware of the generation scheme and can thus register a few AGDs before they are requested by the bots.
The bots, however, query all generated domains one after the other, most of which lead to non-existing domain (NXD) responses, until they find a registered domain that enables communication with their C2 server.
Through this process, the botnet master forces defenders into an asymmetrical position.
Unlike hard-coded IP addresses or domains, which can be easily blocked once identified, defenders must block each AGD to ensure success.
In contrast, an attacker only needs to register a few AGDs that have not been blocked in order to maintain control of the botnet.

In light of the ever-increasing number of Internet-connected devices, detecting bots is becoming increasingly important.
In the past, researchers have devised various methods to contain the threat posed by botnets.
One promising approach is to monitor DNS traffic to identify malware-infected hosts and block communication with C2 servers by detecting malicious AGDs.
Proposed state-of-the-art classifiers are based on deep learning (DL) and achieve true-positive rates (TPRs) of 99.98\% at false-positive rates (FPRs) of 0.15\%~\cite{drichel_analyzing_2020}.
Although these results sound promising, the robustness of the classifiers has not been adequately addressed in related work, which is particularly dangerous since the classifiers operate in an adversarial environment where an attacker is actively trying~to~avoid detection.

Recently, machine learning (ML) explainability techniques were used to uncover five biases in the training data of state-of-the-art DGA classifiers that attackers can exploit to evade detection~\cite{drichel_false_2023}.
For instance, an attacker can simply replace the top-level domain (TLD) or prepend \textit{www.} to a malicious AGD, that would otherwise be detected with high confidence, to evade detection.
The TLD and all domain labels preceding the effective second-level domain (e2LD)\footnote{The e2LD corresponds to the part of a domain that must be registered by a botnet master or a benign party via any TLDs, public suffixes, or dynamic DNS service~providers.} should not have an influence on the prediction of the classifier.
By training on e2LDs, the authors of~\cite{drichel_false_2023} derive a classifier that is significantly less affected by the identified biases.
However, this is accompanied by an enormous degradation in classification performance, which is mitigated by combining the bias-reduced classifier with a classifier trained on fully qualified domain names (FQDNs) into a single classification system.
\copyrightnotice{\copyright\space Copyright held by the owner/author(s) 2024. This is the author's version of the work. It is posted here for your personal use. Not for redistribution. The definitive version was published in ACM Asia Conference on Computer and Communications Security (ASIA CCS ’24), https://doi.org/10.1145/3634737.3656287} 

In addition to training biases, DL-based classifiers are known to be susceptible to adversarial examples (AEs), i.e., carefully crafted input perturbations that cause a classifier to misclassify~\cite{Szegedy2013}.
Related work has only partially addressed the robustness of DGA classifiers against AEs and the attacks that generate them (adversarial attacks).
Most works only propose attacks based on character perturbations or generative adversarial networks (GANs), demonstrate their effectiveness against unhardened DGA classifiers, and finally quantify the increased robustness of the models after adversarial training (AT), i.e., after retraining the classifier using a dataset that additionally contains domain names generated by the proposed attack.
Thereby, related work leaves out the bigger picture of the general robustness of DGA classifiers against known but also against yet unknown attacks, which is particularly important in practice.
Specifically related work~\cite{Zhai2022,Peck2019,Liu2021,Spooren2019,Anderson2016,Corley2019,Gould2020,Shu2022,Yun2019,Liu2022,Sidi2019,nie2022pkdga,Hu2023,Zheng2021} falls short on the following:

First, the threat model is inadequately defined.
Adversarial attacks have been extensively and predominantly studied for computer vision models~\cite{Akhtar2018,Akhtar2021}.
There, however, the image classifiers operate in a different environment, and the adversarial attacks focus on generating AEs that are similar to the unperturbed input.
The advantage of sticking to similarity is that AEs can fool an image classifier while the input noise is imperceptible to the human eye.
In contrast, domains queried by an undetected bot are typically not manually inspected by a human, thus bounding the perturbations is not necessary.
Focusing only on small input perturbations can instead have a negative impact and provide a false sense of security, as classifiers hardened with these AEs could still be fooled by an adversary performing the same attack with an increased bound.

Second, the robustness evaluations performed in related work are inadequate and do not come close to a real-world setting.
In related work, either only new attacks are proposed and the vulnerability of classifiers to them is shown, or when classifiers are hardened, only a single attack is used for AT and robustness is evaluated only against the same attack.
The few works that also evaluate robustness against attacks that were not used to harden the classifiers use a single black-box attack for AT and evaluate robustness against only one or two other attacks, which is not sufficient to derive statements about the general robustness of classifiers.
In practice, a classifier would be hardened against all known attacks, as it has been shown that models that have been hardened against one attack through AT do not necessarily exhibit improved robustness against other attacks~\cite{Akhtar2021}.
In addition, it is possible that unknown attacks are carried out during the operation of the classifier in practice.
Therefore, the robustness of a classifier should preferably be evaluated against attacks that have not been used to harden it.

Third, most works use artificially generated data based on public top sites rankings such as Tranco~\cite{lepochat_tranco_2019} to train, test, attack, and harden DGA classifiers.
However, artificial data may not accurately reflect the real-world data distribution and therefore lead to bias and misleading results.
Specifically, it has been shown that proposed adversarial attacks against DGA classifiers perform significantly worse on classifiers that were trained on real-world NXD data~\cite{drichel_analyzing_2020}.
In fact, adversarial attacks such as~\cite{Peck2019,Spooren2019} are detected with over 99\% TPR, which completely calls into question the evaluations performed by related work and the degree of robustness achieved by AT using AEs generated by those attacks.

In our work, we address these shortcomings.
To the best of our knowledge, we are the first to systematically and comprehensively perform a critical analysis of the general robustness of state-of-the-art DGA classifiers.
In this context, we implement state-of-the-art adversarial attacks, uncover blind spots in the classifiers, and harden them to significantly increase their robustness. 

In total, we design, implement, and evaluate 32 white-box attacks based on published works from different domains, including image classification, natural language processing (NLP), and adversarial research DGAs.
State-of-the-art gradient-based attacks are not directly applicable to DGA classifiers as the models contain non-differential embedding layers.
Applying these attacks at the embedding level results in adversarial vector representations that must be discretized to adversarial domains to be either used for malicious purposes or to perform AT at the character level.
The problem with the discretization of adversarial vectors is that it must be ensured that the discretized domain is a valid in the sense of RFC1035~\cite{rfc1035} and is still adversarial for the classifier (i.e., the discretized domain is still classified as benign by the classifier).
In this work, we solve this discretization problem by developing and evaluating several controllable discretization algorithms that map adversarial embedding vectors to valid adversarial domains, making a large body of adversarial ML research directly applicable to DGA classification.

In this course, we compare AT on discretized domains with training on adversarial latent space vectors and propose a novel training scheme that leverages both to significantly improve robustness.
We evaluate the robustness of DGA classifiers using a comprehensive leave-one-group-out (LOGO) evaluation, where we train classifiers on all but one attack and evaluate their robustness on the omitted attack.
Thereby, we are close to a real-world setting and are able to quantify the robustness against unknown attacks.

In our work, we focus on the robustness of the proposed bias-reduced DGA classifier~\cite{drichel_false_2023} to derive the most advantageous approach for practical use.
By analyzing the domains generated by our adversarial attacks, we uncover two additional biases inherent in the bias-reduced DGA classifier that can be easily exploited.
We thus show that it is not sufficient to rely only on explainability techniques to identify and remove biases inherent in state-of-the-art classifiers, and propose to complement such analysis with AT.

\section{Preliminaries \&  Related Work}
\label{sec:preliminaries_rw}
In this section, we first introduce the fundamentals that are helpful for understanding and categorizing the rest of the work.
Then, we specifically focus on related work dealing with ML for DGA detection, as well as adversarial ML in related fields.

\subsection{Adversarial Attacks}
ML systems are susceptible to AEs designed to mislead the system.
Examples include images altered to fool image classifiers~\cite{Szegedy2013}, hoodies designed to make their wearer become invisible to object detectors~\cite{Wu2020}, and malware binaries tailored to bypass malware detectors~\cite{Lucas2023}.
Szegedy et al.~\cite{Szegedy2013} were among the first to study AEs for neural networks (NNs).
They found that NN predictions can be effectively manipulated by applying small malicious perturbations to its inputs.
Since 2014, well over 8000 papers have been published concerning themselves with the threat of AEs~\cite{Carlini2019a}.
Some of these works focus on creating novel attacks (e.g.,~\cite{Carlini2017a,Croce2020,Ebrahimi2017,Goodfellow2014,Madry2017}), while others try to defend NNs against AEs (e.g.,~\cite{Croce2020a,Guo2018,Madry2017,Yang2022}).
AEs are predominantly studied for image classifiers.
One intriguing property of AEs is their frequently observed ability to transfer to other models trained to solve the same task \cite{Liu2016,Papernot2016b}.
Over time, the concept of AEs has been transferred to many other domains, including NLP and DGA classification.

Adversarial attacks are algorithms that generate AEs.
It is well known that gradient-based white-box attacks are among the most potent attacks and should always be stronger than black-box attacks~\cite{Carlini2019}.
Therefore, in this work we focus on white-box attacks.

\subsection{Defenses against Adversarial Attacks}
Defenses against AEs can be divided into four categories~\cite{Akhtar2018,Akhtar2021}:

\emph{Model-Robustification} defenses alter the model to make it more robust against perturbation attacks.
AT~\cite{Madry2017} defenses are among the most well-known defenses of this class~\cite{Akhtar2021}.
These defenses work by incorporating AEs in the training process of the network. 

\emph{Perturbation-Removal} defenses modify the input before passing it on to the network, trying to remove adversarial perturbations.
For instance, Pruthi et al.~\cite{Pruthi2019} proposed one such defense to protect NLP classifiers by placing a word recognition model before the targeted classifier defending against adversarial spelling mistakes.

\emph{External-Module} defenses use additional modules besides the classification network to deal with AEs.
AE detectors that try to detect the presence of AEs are well-known examples of this kind of defense (e.g., ContraNet~\cite{Yang2022}).

In contrast to these defenses, which are typically only validated empirically, \emph{Certified} defenses provide provable robustness guarantees.
These defenses commonly work by training a model that is as robust as possible and then using an algorithm to generate a robustness certificate.
Recent studies~\cite{Li2023} show that there is significant progress being made in this area.

Defending against AEs is challenging: Many proposed defenses get circumvented with more potent attacks~\cite{Athalye2018,Carlini2017,Tramer2020,Zimmermann2022}.
Athalye et al.~\cite{Athalye2018} analyzed nine defenses and found that many defenses rely on gradient masking, which decreases the usefulness of the information provided by local gradients.
Some techniques that can be used to circumvent gradient masking are random restarts and testing against AEs generated on the unhardened model~\cite{Carlini2019}.
AT is one technique that stood the test of time~\cite{Zimmermann2022} and is generally considered beneficial in many cases.
Projected gradient descent (PGD)~\cite{Madry2017} based AT has become the de-facto standard method for increasing the robust accuracy of a model in practice~\cite{Bai2021}. 
Therefore, in this work, we focus on AT to harden DGA classifiers.


\subsection{Domain Generation Algorithm Classifiers}
Several approaches have been devised in the past to detect DGA activities.
Some approaches are designed to only notice the presence of AGDs (binary classification, e.g.,~\cite{drichel_analyzing_2020,schuppen_fanci_2018, woodbridge_predicting_2016,yu_character_2018}), while others additionally try to predict the malware family the AGDs belongs to (multiclass classification, e.g.,~\cite{drichel_analyzing_2020,drichel_first_2021,tran_lstm_2018,woodbridge_predicting_2016}).

Further, approaches can be divided into context-less (e.g.,~\cite{drichel_analyzing_2020,saxe_expose_2017,schuppen_fanci_2018,woodbridge_predicting_2016,yu_character_2018,tran_lstm_2018,drichel_false_2023}) and context-aware (e.g.,~\cite{antonakakis_throwaway_2012,bilge_exposure_2014,shi_malicious_2018,grill_detecting_2015,schiavoni_phoenix_2014,yadav_winning_2012, antonakakis_detecting_2011,antonakakis_building_2010}) classification methods:
Approaches that base their decision solely on information that can be extracted from a single domain name are called context-less.
Context-aware approaches leverage additional contextual information aiming at improving classification performance.
In the past, several studies~\cite{drichel_analyzing_2020,schuppen_fanci_2018,woodbridge_predicting_2016,yu_character_2018} have concluded that context-less approaches achieve similar or even better classification performance compared to context-aware systems, while they have higher throughput and are less invasive to user privacy.

The classifiers based on ML within the group of context-less approaches can be further subdivided into feature-based  (e.g., support vector machines or random forests~\cite{schuppen_fanci_2018,drichel_first_2021,bilge_exposure_2014}) and feature-less (DL, e.g., recurrent or convolutional neural networks~\cite{drichel_analyzing_2020,woodbridge_predicting_2016,yu_character_2018,saxe_expose_2017}) classifiers.
Feature-based classifiers require prior feature engineering by domain experts, which is an extensive and manual task before they can be trained.
DL-based approaches, on the other hand, learn by themselves to extract the relevant features for classification.
Previous studies~\cite{drichel_analyzing_2020,woodbridge_predicting_2016,Peck2019,Spooren2019,sivaguru_evaluation_2018} have shown that DL-based approaches achieve better classification performance compared to feature-based classifiers.

Since most AGDs are unregistered and therefore most bot queries result in NXD responses, several approaches (e.g.,~\cite{schuppen_fanci_2018,drichel_first_2021,drichel_analyzing_2020,drichel_making_2020,antonakakis_throwaway_2012,yadav_winning_2012,tong_far_2020}) focus on classifying non-resolving DNS traffic (NX-traffic) to detect botnet activity.
Focusing on NX-traffic has the added benefit of being easier to monitor since the amount of NX-traffic is an order of magnitude smaller than the amount of full DNS traffic.
Additionally, it is less privacy sensitive because NXDs typically do not contain user-entered domains except for typo domains, and botnet activity can usually be detected before the bots are demanded to perform malicious actions.
Furthermore, proposed attacks on DGA classifiers have been shown to perform worse on classifiers trained on NXDs than on classifiers trained only on resolving domains~\cite{drichel_analyzing_2020}.

Recently, the authors of~\cite{drichel_false_2023} analyzed the context-less DL-based DGA classifiers and revealed several biases that strongly influence the classifiers' predictions without being closely related to the underlying problem and can be exploited by an adversary to easily bypass detection.
To mitigate this issue, the authors propose a bias-reduced classifier that is trained only on e2LDs.

In this work, we acknowledge the extensive previous efforts carried out in related work by following their recommendations.
In particular, we focus on bias-reduced DGA classifiers trained on e2LDs extracted from real-world NX-traffic to evaluate and improve their robustness against adversarial attacks.

\subsection{Adversarial DGAs}
Several previous works have concerned themselves with the robustness of DGA classifiers and the creation of adversarial DGAs that aim to evade detection.
Table~\ref{tab:comparison} summarizes the results of these studies and places our work in context with them.

\rowcolors{3}{white}{gray!20}
\begin{table*}[t]
	\centering
	\caption{Comparison of related work dealing with adversarial machine learning for DGA classification.}
	\scriptsize
	\resizebox{\textwidth}{!}{
	\begin{threeparttable}
	\begin{tabular}{l|l|l|c|c|c|c|cl}
		 & & & \textbf{Avail}- & \textbf{Threat} & \textbf{Real} & \textbf{Number} &  & \\
		 \multirow{-2}{*}{\textbf{Approach}} & \multirow{-2}{*}{\textbf{Year}} & \multirow{-2}{*}{\textbf{AE generation method}} & \textbf{ability} & \textbf{Model} & \textbf{Data} & \textbf{Attacks} & \multicolumn{2}{l}{\multirow{-2}{*}{\textbf{Robustness Evaluation}}} \\
		\hline \Tstrut

		CDGA~\cite{Zhai2022} & 2022 & Generative Network (WGAN-GP) & \tabbad & \tabbad & \tabbad & 4 & \tabbad & No robustness evaluation against attacks\\ 
		CharBot~\cite{Peck2019} & 2019 & Character Perturbations of benign domains & \tabgod  & \tabbad & \tabgod & 3 & \tabmed  & 3 $\cdot$ (train on 1 attack, test on 2 unknown attacks) \\
		CLETer~\cite{Liu2021} & 2021 & Character Perturbations of AGDs & \tabbad & \tabbad & \tabbad & 2 & \tabbad & Adversarial training, no tests on unknown attacks\\ 
		DeceptionDGA~\cite{Spooren2019} & 2019 & Iterative Feature-Engineering & \tabmed \tnote{1} & \tabbad & \tabbad & 1 & \tabbad & No robustness evaluation against attacks \\ 
		DeepDGA~\cite{Anderson2016} & 2016 & Generative Network (GAN) & \tabmed \tnote{1} & \tabbad & \tabbad & 1 & \tabbad & No robustness evaluation against attacks \\
		DomainGAN~\cite{Corley2019} & 2019 & Generative Network (GAN/LSGAN/WGAN-GP) & \tabbad & \tabbad & \tabbad & 3 & \tabbad & Adversarial training, no tests on unknown attacks\\ 
		Gould et al.~\cite{Gould2020} & 2020 & Generative Network (WGAN-GP) &  \tabbad &  \tabbad &  \tabbad &  1 & \tabbad & No robustness evaluation against attacks \\ 
		GWDGA~\cite{Shu2022} & 2022 & Generative Network (VAE) & \tabbad & \tabbad & \tabbad & 2 &  \tabbad & Adversarial training, no tests on unknown attacks\\
		KhaosDGA~\cite{Yun2019} & 2019 & Generative Network (WGAN-GP) & \tabmed \tnote{1} & \tabbad & \tabbad & 2 & \tabmed & 2 $\cdot$ (train on 1 attack, test on 1 unknown attack)\\
		Liu et al.~\cite{Liu2022} & 2022 & Generative Network (GAN) &  \tabbad &  \tabbad &  \tabbad &  3 & \tabbad & No robustness evaluation against attacks \\ 
		MaskDGA~\cite{Sidi2019} & 2019 & Character Perturbations of AGDs & \tabgod \tnote{1} & \tabgod & \tabgod & 3 & \tabmed & 3 $\cdot$ (train on 1 attack, test on 2 unknown attacks)\\
		PKDGA~\cite{nie2022pkdga} & 2022 & Generative Network (GAN) & \tabgod & \tabbad & \tabbad & 2 & \tabmed & 2 $\cdot$ (train on 1 attack, test on 1 unknown attack) \\
		ReplaceDGA~\cite{Hu2023} & 2023 & Character Perturbations of benign domains & \tabbad &  \tabbad &  \tabbad &  7 & \tabbad & Adversarial training, no tests on unknown attacks \\ 
		ShadowDGA~\cite{Zheng2021} & 2021 & Generative Network (GAN) & \tabbad & \tabgod & \tabbad &  2 & \tabmed & 2 $\cdot$ (train on 1 attack, test on 1 unknown attack)\\
		\hline \Tstrut
		Our Work & 2024 & 32 white- \& 4 black-box (\cite{Spooren2019,Anderson2016,Yun2019,Sidi2019}) attacks &  \tabgod &  \tabgod &  \tabgod &  36 & \tabgod & Comprehensive leave-one-attack-out study\\
	\end{tabular}
	\begin{tablenotes}
			\item[1] Pre-computed samples are available.
		\end{tablenotes}
	\end{threeparttable}
	}
	\label{tab:comparison}
\end{table*}

Most works use either generative networks based on different types of GANs or calculated character perturbations to generate AEs.
One exception is DeceptionDGA~\cite{Spooren2019}, an iteratively developed algorithm that creates AGDs that aim to exploit the manually engineered features of the feature-based classifier FANCI~\cite{schuppen_fanci_2018}.

As shown in Table~\ref{tab:comparison}, the source code is only publicly available for 3 out of 14 works, for 4 out of 14 works pre-calculated AEs can be found on the Internet.
The limited availability considerably restricts the possibility of evaluating all attacks in a standardized study and training a classifier that is as robust as possible for practical use.

As far as a well-defined threat model is concerned, only 2 out of 14 papers mention a threat model at all (see Table~\ref{tab:comparison}), which is not necessarily designed to derive the most robust classifier possible.

Furthermore, only 2 out of 14 papers use real-world data for their robustness assessment.
Thus, 12 out of 14 works use artificial data, which may not accurately reflect the real-world data distribution and therefore lead to bias and misleading results.

In each of the 14 works, between 1 and 7 adversarial attacks are examined to varying degrees (see Table~\ref{tab:comparison}).
When it comes to robustness analyses, 5 out of 14 works demonstrate only the vulnerability to adversarial attacks and do not perform any hardening.
4 out of 14 works perform AT and only evaluate the robustness of the classifiers against attacks that were used to harden the classifier.
And 5 out of 14 works perform AT on a single attack and evaluate the classifiers' robustness against only one or two other attacks.
These procedures are, however, not sufficient to derive statements about the general robustness of DGA classifiers.

In this paper, we address the shortcomings of related work.
We implement 32 state-of-the-art white-box attacks for the DGA detection use case, which we use in a comprehensive LOGO study to evaluate the overall robustness of bias-reduced DGA classifiers on real-world data.
Additionally, we evaluate the robustness of the classifier against four proposed black-box attacks~(\cite{Spooren2019,Anderson2016,Yun2019,Sidi2019}) and operate in a classification environment that is close to the real world, allowing us to assess the classifier's robustness against unknown attacks.
Finally, we define our threat model with the aim of deriving the most robust classifier.

\subsection{Adversarial ML in Related Domains}
The domain of raw-binary malware classification is conceptually very similar to the domain of DGA classification:
Classifiers operate in an environment directly facing adversaries in both domains.
Furthermore, classifiers of both domains operate on discrete inputs that must fulfill strict requirements to be considered valid.
Kreuk et al.~\cite{Kreuk2018} introduce a white-box attack that appends bytes to a malware binary to make it appear benign to a classifier.
They generate the bytes with an iterative version of the FGSM~\cite{Goodfellow2014} attack and reconstruct a discrete input by rounding to the nearest embedding vector using the $L_2$ norm.
Lucas et al.~\cite{Lucas2023} adversarially trained a raw-binary malware classifier on multiple attacks and thereby were able to significantly improve robustness.

NLP classifiers (e.g., such used for sentient analysis) are also related to DGA classifiers as they operate on textual input.
Attacks for NLP models are often designed to generate outputs that are semantically equivalent to the corresponding input (e.g.,~\cite{Jin2020,Li2019,Ren2019,Wang2020,Zang2019}).
From the perspective of generating adversarial domains, this is not required, as end users do not see DNS requests.
Unlike these attacks, several character-substitution attacks only focus on generating strong AEs without regard to semantic equivalence.
HotFlip~\cite{Ebrahimi2017} performs a beam search guided by the gradients of the network to generate AEs.
Yoo et al.~\cite{Yoo2020} compared different search methods for finding discrete NLP AEs and concluded that beam search optimization is one of the most effective approaches.

In our work, we also utilize rounding to the nearest embedding vector using different distance metrics to construct discrete domains.
Further, we harden DGA classifiers using multiple attacks and also include HotFlip as a representative of a beam-search guided attack.

\section{Threat Model}
\label{sec:threat_model}
A well-defined threat model is critical for evaluating attacks and defenses for DGA classifiers.
As DGA classifiers are security-relevant, we argue that similar to Kerckhoff's principle in cryptography, DGA classifiers should be secure even if the attacker knows the system.
Therefore, we assume a white-box scenario in which the adversary has complete knowledge about the system, including all model weights and parameters.
Furthermore, it is commonly assumed that adversaries have read-only access to the system.
This implies that an attacker is unable to alter the deployed model (e.g., through bit-flips), alter the training data (e.g., through data poisoning), or influence the model's prediction in any other way than through changing the input. 
Using this threat model, unhardened and hardened DGA classifiers face the strongest possible adversary (white-box adversary), which means that the adversary has access to all model gradients, can adapt attacks, and directly craft AEs on the target model.

In other domains, such as the domain of image classification, many threat models include bounds or limitations to adversarial perturbations to make the AE indistinguishable from the benign original~\cite{Akhtar2021}.
It does not make sense to impose such limitations here, as DGA classifiers operate on DNS requests, which regular users do not see.

\subsection{Attacker Goals}
We assume that an attacker is the operator of a botnet who already has some algorithm for generating AGDs and wants to apply a transformation to the generated domains to make them appear benign to the classifier. 
Furthermore, as many of the resulting domains as possible should be useable.
A domain is useable by an adversary if it can be registered and has not been previously generated by the transformation function.
To this end, we define the following goals that transformation functions (attacks) should strive to fulfill:
\begin{enumerate}
	\item The resulting domains should fool the DGA classifier.
	\item The resulting domains must be syntactically valid and comply with common rules imposed by registrars to ensure that they can be registered across a wide range of registrars.
	\item The resulting domains should be unregistered (or, in the case of e2LDs, be unregistered across many TLDs).
	\item The transformation function should be injective or have little duplication. 
\end{enumerate}
We design and evaluate all of our attacks with these goals in mind.

\section{Evaluation Setup}
\label{sec:evaluation_setup}
In the following, we present our evaluation setup, including the dataset used, the attacks and training procedures investigated for classifier hardening, and our ethical considerations.

\subsection{Data}
We use one source of malicious labeled data and two distinct sources of benign labeled data for our evaluation.

\subsubsection{Malicious Data: DGArchive}
DGArchive~\cite{Plohmann2016} is a continuously updated collection of AGDs generated by reverse-engineered DGAs.
At the time of writing, DGArchive is the largest open source intelligence feed for AGDs and contains 136 million unique samples generated by 111 different DGAs.

\subsubsection{Benign Data: University Network}
Our first source of benign labeled domains is the central DNS resolver of the RWTH Aachen University campus network, which serves several academic and administrative networks, student residence networks, and the network of the affiliated university hospital.
For our evaluation we select a one-month recording of NXDs from mid-October 2017 to mid-November 2017 which contains approximately 35 million unique NXDs.
Note, we intentionally chose an older NX-traffic recording in order to be able to evaluate whether hardened classifiers generalize well to new environments and are time-robust.
To this end, we use our second source of benign data: the company~network.

\subsubsection{Benign Data: Company Network}
Benign data from this source is only used in a final real-world evaluation study (see Section~\ref{sec:evaluation_rw}).
We extract benign NXDs from several central DNS resolvers of Siemens AG that cover the regions Asia, Europe, and USA.
The data obtained from this source is particularly diverse, allowing us to evaluate the generalization capabilities of the hardened classifiers.
From this source, we obtain a one-month recording of benign NXDs from April 2019, which contains approximately 373 million NXDs. 

\subsubsection{Dataset}
\label{sec:ds}

We follow the recommendations for data preprocessing and dataset creation proposed in~\cite{drichel_false_2023} to train bias-reduced classifiers.
Focusing on e2LDs significantly reduces the number of available unique benign samples from 35 million unique FQDNs to approximately 125k e2LDs.
To create a balanced dataset, we randomly select at most 2350 e2LDs for each DGA in DGArchive, with the constraint that these samples were generated by DGAs before the end of the university data recording period.
By enforcing this constraint, we reduce temporal experimental bias~\cite{pendlebury_tesseract_2019}.
We combine the selected malicious labeled samples with the benign e2LDs from the university network to form a balanced dataset.
Thereby, the dataset consists of approximately 250k samples, where the malicious labeled samples are generated by 84 different DGAs.
The 27 DGAs that were not selected during this process because they emerged at a later stage are used in a real-world study to evaluate the classifiers' ability of detecting unknown DGAs (see Section~\ref{sec:evaluation_rw_unknown}).

\subsection{DGA Classifiers}
In this work, we focus on hardening the recently proposed bias-reduced DGA classifier~\cite{drichel_false_2023} that is based on a residual neural network (ResNet).\footnote{While our robustness study focuses on the bias-reduced ResNet model, we also examined other models based on different architectures, including the Long Short-Term Memory model~\cite{woodbridge_predicting_2016} and the convolutional NN~\cite{yu_character_2018} that includes two stacked convolutional layers, and assessed similar vulnerability and robustness~properties.}
To make our results more reliable, we train five classifiers using a stratified five-fold cross-validation approach.
Thereby, each fold has 75\% of the data in its training set, 5\% of the data in its validation set, and the remaining 20\% of the data in its testing set.
We use the validation set to decide how many epochs to train using early stopping with a patience of five epochs.

When we attack the classifiers, we generate our AEs based on the AGDs in the test set of each fold.
Thereby, we can guarantee that no information about the samples used for an attack is included during the training of a classifier.
Otherwise, we might overestimate the robustness.
When comparing the effectiveness of different attacks, we use the false-negative rates (FNRs), as this is the most important metric that an attacker needs to maximize.
Unless otherwise stated, all metrics are given as average values across all five folds.

\subsection{Adversarial Attacks}
In this paper, we distinguish between embedding-space attacks, which generate adversarial latent space vectors, and discrete attacks, which yield discrete adversarial domains.

\subsubsection{Embedding-Space Attacks}
It is well established that gradient-based white-box attacks are the most powerful attacks against DL classifiers~\cite{Carlini2019}. 
However, proposed state-of-the-art attacks cannot be directly applied to DGA classification models since a non-differentiable embedding layer is used.
Nevertheless, we can apply these attacks to the layers following the embedding and use a discretization algorithm to extract adversarial domains.

Evaluating the robustness of a DL model is challenging, and choosing the right attacks is essential to prevent overestimating robustness~\cite{Carlini2019}.
After a thorough review of recent literature, we have decided to include the following attacks in our evaluation:
Projected Gradient Descent (PGD)~\cite{Madry2017} both in its $L_2$ and $L_\infty$ variants, Carlini and Wagner (C\&W)~\cite{Carlini2017a} in its $L_2$ variant, and AutoAttack~\cite{Croce2020} in its $L_2$ and $L_\infty$ variants, which we adapt to binary classifiers (BAT).
The Appendix~\ref{sec:appendix_embedding_space_attacks} details the selected attacks and the hyperparameter ranges examined in more detail.

\subsubsection{Discretization}
The adversarial attacks generate adversarial embedding vectors $v^{adv} \in \mathbb{R}^{d \times n}$, where $d$ is the embedding dimension and $n$ is the input length. Adversaries need registerable e2LDs $w \in \Sigma^{n}$ (where $\Sigma$ is the alphabet) whose embedding $\mathrm{emb}(w) \in \mathbb{R}^{d \times n}$  passes the detector. However, it is almost certain that $\mathrm{emb}^{-1}(v^{adv}) = \emptyset$, which makes the raw adversarial embedding vector useless to an adversary without additional post-processing.
We propose to use a character-discretization function $g_{\Sigma, W, D}$ that maps an embedding vector to its nearest neighbor in the alphabet~$\Sigma$:
\begin{equation*}
	g_{\Sigma, W, D}: \mathbb{R}^{d} \rightarrow \Sigma \qquad x \mapsto \argmin_{c \in \Sigma} D(W_c, x),
\end{equation*}
where $W_c$ is the embedding of a character $c \in \Sigma$.
Distances are calculated based on the normalized embedding matrix $W \in \mathbb{R}^{\lvert \Sigma \rvert \times d}$ in combination with a distance measure $D$. 
In this work, we evaluate $L_2$, $L_\infty$, and cosine distance ($D_c$) based rounding.

This discretization function can be applied character-wise to an adversarial embedding vector $v^{adv}$ to retrieve an adversarial domain.
However, one of the challenges with this approach is that the padding character ($\sim$) may appear at arbitrary positions in the rounded domain.
One approach would be to disregard all characters following the first occurrence of the padding character.
However, many of the domains discretized in this way are still syntactically invalid.
A valid e2LD is a maximum of 63 characters long and consists of ASCII letters, digits, and the hyphen symbol with the additional restriction that the hyphen must not be at the beginning or end of the label~\cite{rfc5890}.
Unicode characters may be embedded in these labels using an ASCII-compatible encoding.
Such internationalized domain names (IDNs) start with \nolinkurl{xn--}.
Registrars supporting IDNs are advised to disallow registrations of domains with hyphens in both the third and fourth position of a domain unless that domain is a valid A-label as described in RFC5890~\cite{rfc5890}.

To tackle this issue, we devise a controllable discretization scheme that maps an arbitrary embedding vector to a valid e2LD of any pre-determined length.
Given a distance metric $D$, the alphabet of valid e2LD characters $\Sigma$, an adversarial embedding vector $v^{adv}$, and a pre-determined output length $\ell$, we define for each output character $i$:

\rowcolors{1}{}{}
\begin{equation*}
	\mathrm{disc}_{\Sigma, W, D}(v^{adv}, \ell)_i =
	\begin{cases}
		g_{\Sigma \setminus \{\sim, \text{-}\}, W, D}(v^{adv}_i) & \text{if } i = 1 \lor i = \ell \\
		g_{\Sigma \setminus \{\sim \}, W, D}(v^{adv}_i) & \text{if } 1 < i < \ell \\
		\sim & \text{if } i > \ell \\
	\end{cases}
\end{equation*}

By pre-determining the length of the domain, we can restrict the rounding mechanism to not round to padding characters in the middle of the domain, to not round to a hyphen character at the start and end of a domain, and to not round to anything except the padding character outside of the domain area. 
We must also incorporate the constraint that the third and fourth characters cannot both be a hyphen. We do so using the following conflict resolution scheme: If both the third and fourth characters resolve to a hyphen, the one closer to the hyphen becomes the hyphen, and the other becomes the second-closest character. 

To build a full discretization algorithm, we need to combine the discretization scheme with another algorithm to determine the output domain's length.

\paragraph{Length-Cutoff (LCO)}
One approach of choosing the domain length is by taking the index of the first character embedding vector whose nearest neighbor is the padding character.
This is similar to dropping all characters after the first occurrence of the padding character.
The domain names generated using this algorithm are, however, often very short.
This could be a problem for attackers as very short names tend to already be registered.
Therefore, we decide to incorporate a minimum length of seven characters for domain names. 
We refer to this approach as \emph{Length-Cutoff (LCO)} and test it with the discussed attacks and rounding norms.

We decided on a minimum length of seven characters based on a worst-case analysis:
The Domain Name Industry Brief (DNIB) from Q3 2023~\cite{dnibQ3-2023} reports that the TLD with the most registrations is \textit{.com} with a total of $160.8$ million domains.
If all those domains were six characters long, roughly $6.6 \%$ of all available six-character domains would not be available.
Repeating the same experiment with seven characters shows that at most $0.2 \%$ of all seven-character domains would not be available.
Hence, the prescribed minimum length of seven characters significantly reduces the likelihood of already registered domains being generated. 

These thoughts lead to the following domain-length formula:
\begin{equation*}
	l(v^{adv}) = \max \left(7 \;,\; \min \{ \ 1 \leq i < 63 \ | \ g_{\Sigma, W, D}(v^{adv}_{i+1}) = \ \sim\ \} \cup \{ 63 \}\right)
\end{equation*}

\paragraph{Length Brute-Force (LBF)}
Selecting the length of the output domain name by the occurrence of the first character whose nearest neighbor is the padding character comes with the downside that we loose all information encoded in all characters following it.
This is suboptimal since our adversarial attacks perform alterations along the entire domain length.
As e2LDs are at most 63 characters long, performing a brute-force attack to find the optimal length is feasible.
Therefore, we implement a different algorithm that chooses the domain length as the one that maximizes the loss $L$ of the classifier $f$ and the true label $y$.
Formally, we choose the length as:

\begin{equation*}
	l(v^{adv}) = \argmax_{7 \leq i \leq 63} L(f(\mathrm{disc}_{\Sigma, W, D}(v^{adv}, i)), y)
\end{equation*}
This attack is strictly stronger than the previous attack. 
We call this approach \emph{Length Brute-Force (LBF)} and test it with the discussed attacks and rounding schemes.

\subsubsection{Discrete Attacks}
In this section, we introduce discrete attacks that directly work with textual input and thus do not require an additional discretization step.
After a thorough literature review, we find that many attacks rely on the principle of replacing characters or words based on the saliency of the element measured by the magnitude of the gradients (e.g.,~\cite{Ebrahimi2017,Li2019,Ren2019}).
Many of these attacks impose further restrictions that try to hide the perturbations from humans by, e.g., replacing characters with ones that appear similar~\cite{Li2019} or replacing words with synonyms~\cite{Ren2019}.
Such mechanisms are not required for crafting adversarial domains as DNS requests are not seen by users.
Yoo et al.~\cite{Yoo2020} found that attacks using a beam search are among the most powerful attacks.
Therefore, we choose to implement HotFlip~\cite{Ebrahimi2017} as a representative of a white-box beam-search character-swap attack.

Apart from attacks designed for general-purpose NLP models, there are also attacks tailored to the task of evading DGA classifiers.
These attacks are often designed as black-box (BB) adversarial DGAs.
MaskDGA~\cite{Sidi2019} also appears to be a black-box attack on the outside.
However, internally, the MaskDGA attack is comprised of a white-box (WB) character-swap attack that is applied to a hidden substitute model.
The attack is abusing the transferability of AEs to fool other classifiers. 
We therefore choose to implement MaskDGA as a representative of a white-box attack that aims at evading DGA classifiers. We refer to this attack as MaskDGA-WB.

In addition, we evaluate the classifiers against pre-generated samples of DeceptionDGA~\cite{Spooren2019}, DeepDGA~\cite{Anderson2016}, KhaosDGA~\cite{Yun2019}, and MaskDGA~\cite{Sidi2019}. 
We denote the samples generated by these black-box attacks with the suffix \textit{-BB}.

\subsection{Adversarial Training Schemes}
AT revolves around incorporating AEs in the training loop of NNs.
In this work, we distinguish between training on adversarial latent space vectors (e.g.~\cite{Zhu2019}), and training on discrete AEs (e.g.~\cite{Yoo2021}).
On the one hand, an adversary that can generate adversarial embedding vectors is strictly more powerful than an adversary that can only generate adversarial domains (as not all embedding vectors are reachable).
On the other hand, training on adversarial embedding vectors may waste model capacity for precisely this reason.

During AT, we observe that early stopping does not work well, as the validation loss tends to increase for the first epochs before decreasing again, leading early stopping to terminate too early.
The analysis of the log files of our experimental runs shows that training for a maximum of $50$ epochs generally works well.

\subsubsection{Embedding-Space Adversarial Training}
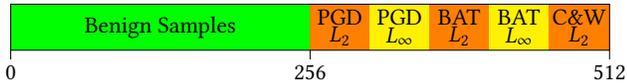
\begin{figure}
	\centering
 	\resizebox{\columnwidth}{!}{
	\begin{tikzpicture}
		\def\unitwidth{13/512}
		
		\node[draw=none] (benign-samples) at ($\unitwidth*(256,0)$) {};
		\fill [fill=green] (0, 0) rectangle ($(benign-samples) + (0,1)$) node[midway] {\huge Benign Samples};
		
		\xdef\lastnode{benign-samples}
		
		\foreach \attack/\width/\attackColor/\attackCaption in {
			atl2/51/orange/PGD\\\huge $L_2$,
			atlinf/51/yellow/PGD\\\huge$L_\infty$,
			pgdl2/51/orange/BAT\\\huge$L_2$,
			pgdlinf/51/yellow/BAT\\\huge$L_\infty$,
			cwl2/52/orange/C\&W\\\huge$L_2$
		} {
			\node[draw=none] (\attack) at ($(\lastnode) + \unitwidth*(\width,0)$) {};
			\fill [fill=\attackColor] ($(\lastnode)$) rectangle ($(\attack) + (0,1)$) node[midway,align=center] {\huge \attackCaption};
			\xdef\lastnode{\attack}
		}
		
		\draw (0, 0) rectangle ($\unitwidth*(512,0) + (0, 1)$);
		
		\foreach \tick in {0, 256, 512} {
			\draw ($\unitwidth*(\tick,0)$) -- ++(0, -0.2) node[below] {\huge $\tick$};
		}

	\end{tikzpicture}
	}
	\caption{Minibatch layout of embedding-space AT.}
	\label{fig:EmbeddingSpaceATMinibatch}
\end{figure}

We train against all five embedding-space attacks.
To this end, we generate an equal number of samples per attack for every batch.
Fig.~\ref{fig:EmbeddingSpaceATMinibatch} shows the composition of the minibatches in this AT setting.

In addition to allocating the samples to different attacks, we must consider how we choose the attack hyperparameters.
For all attacks (except C\&W), the most important hyperparameter is the perturbation budget. 
As our threat model allows for unbounded perturbations, it is tempting to select the attack hyperparameters accordingly.
Doing so is, however, suboptimal as many of our attacks initialize themselves to a random point in the allowed perturbation space.
This would mostly remove the connection to the training data in an unbounded setting.
Furthermore, Kurakin et al.~\cite{Kurakin2017} observed that training against one pre-set perturbation budget does not improve robustness against other perturbation budgets to the same degree.
Therefore, we randomly select a perturbation budget for the attacks for each minibatch.
For $L_2$ bounded attacks (PGD $L_2$, BAT $L_2$) we sample $\varepsilon \sim U(0.5, \sqrt{63 * 128})$.
For $L_\infty$ bounded attacks (PGD $L_\infty$, BAT $L_\infty$), we sample $\varepsilon \sim U(0.01, 1)$.
The chosen bounds correspond to the minimum and maximum perturbation budgets explained in Appendix~\ref{sec:appendix_embedding_space_attacks}.
C\&W is an optimization technique that aims to find an AE with minimal perturbation.
It, therefore, does not have a perturbation budget hyperparameter.
The most important hyperparameter of the C\&W attack is confidence $\kappa$.
High values of $\kappa$ incentivize the attack to search for an AE that get misclassified with higher confidence.
We started sampling $\kappa \sim U(0, 100)$ but noticed that during AT the attack would no longer be able to find AEs for large values of $\kappa$.
We set $\kappa = 0$ for $50 \%$ of cases to combat~this.

\subsubsection{Discrete Domain Adversarial Training}
\begin{figure}
	\centering
 	\resizebox{\columnwidth}{!}{
	\begin{tikzpicture}
		
		\def\unitwidth{13/512}
		
		\node[draw=none] (benign-samples) at ($\unitwidth*(256,0)$) {};
		\fill [fill=green] (0, 0) rectangle ($(benign-samples) + (0,1)$) node[midway] {\LARGE Benign Samples};
		
		\xdef\lastnode{benign-samples}
		
		\foreach \attack/\width/\attackColor/\attackCaption in {
			atl2/36/orange/PGD\\\LARGE$L_2$,
			atlinf/36/yellow/PGD\\\LARGE$L_\infty$,
			pgdl2/36/orange/BAT\\\LARGE$L_2$,
			pgdlinf/36/yellow/BAT\\\LARGE$L_\infty$,
			cwl2/36/orange/C\&W\\\LARGE$L_2$,
			hotflip/38/violet/\LARGE HF,
			maskdga/38/magenta/\LARGE M\\\LARGE DGA
		} {
			\node[draw=none] (\attack) at ($(\lastnode) + \unitwidth*(\width,0)$) {};
			\fill [fill=\attackColor] ($(\lastnode)$) rectangle ($(\attack) + (0,1)$) node[midway,align=center] {\large \attackCaption};
			\xdef\lastnode{\attack}
		}
		
		\draw (0, 0) rectangle ($\unitwidth*(512,0) + (0, 1)$);
		
		\foreach \tick in {0, 256, 512} {
			\draw ($\unitwidth*(\tick,0)$) -- ++(0, -0.2) node[below] {\huge $\tick$};
		}
		
		\def\unitwidth{5/6}
		\node[draw=none] (start-split) at (4.2,1.5) {};
		
		\draw ($(start-split)$) rectangle ($(start-split) + \unitwidth*(6.6,0) + (0, 1)$);
		
		\xdef\lastnode{start-split}
		
		\foreach \attack/\attackColor/\attackCaption in {
			lbf_l2/orange/LBF\\\LARGE$L_2$,
			lbf_linf/yellow/LBF\\\LARGE$L_\infty$,
			lbf_cos/violet/LBF\\\LARGE$D_c$,
			lco_l2/orange/LCO\\\LARGE$L_2$,
			lco_linf/yellow/LCO\\\LARGE$L_\infty$,
			lco_cos/violet/LCO\\\LARGE$D_c$
		} {
			\node[draw=none] (\attack) at ($(\lastnode) + \unitwidth*(1.1,0)$) {};
			\draw ($(\lastnode)$) rectangle ($(\attack) + (0,1)$) node[midway,align=center] {\LARGE \attackCaption};
			\xdef\lastnode{\attack}
		}
		
		\draw ($(benign-samples)$) -- ++(0, 1);
		\draw ($(cwl2)$) -- ++(0, 1);
		\draw ($(benign-samples) + (0, 1)$) -- ($(start-split)$);
		\draw ($(atl2) + (0,1)$) -- ($(lco_cos)$);

	\end{tikzpicture}
	}
	\caption{Minibatch layout of discrete domain AT.}
	\label{fig:DiscretDomainATMinibatch}
\end{figure}
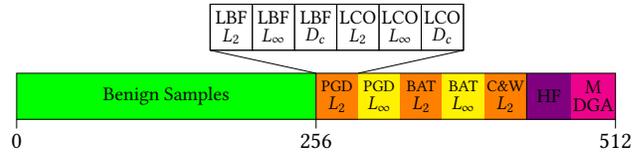

In addition to training on adversarial embedding vectors, we also train on adversarial domains.
We developed 32 different white-box algorithms that can generate adversarial domains:
We can pair all five of our embedding space attacks with all six discretization schemes.
On top of that, we also perform AT against our MaskDGA-WB and HotFlip implementations.
To not under-represent MaskDGA-WB and HotFlip, we split the adversarial batch into seven pieces: The first five belong to the five embedding space attacks, and the last two belong to HotFlip and MaskDGA-WB.
We split each of the embedding space attack batches into six pieces to accommodate the different discretization schemes.
Fig.~\ref{fig:DiscretDomainATMinibatch} visualizes the proportions of different attacks in the training batches for discrete domain AT. 
The discretization schemes and MaskDGA-WB are parameter-free.
The embedding-space attacks use the same hyperparameter configuration as discussed in the previous section.
For HotFlip, we randomly sample the number of flips from $n \sim U(\{1, ..., 10\})$.

\subsubsection{Joint Adversarial Training}
Performing AT on adversarial embedding vectors and adversarial domains together might be beneficial.
In the discrete domain AT setting, we already split the $256$ samples among $32$ attacks.
This results in some attacks only generating $6$ samples.
As there is some initialization overhead for each attack, we decide not to split the batches further.
Instead, we perform joint AT by randomly alternating between training on embedding-space batches and discrete domain batches.

\subsection{Ethical Considerations}
Our institution does not yet have an ethics review board that could have approved this study.
However, unlike several other works that propose adversarial DGAs that can bypass state-of-the-art classifiers, this work addresses robustness and hardening of classifiers.
Although we make the source code of the white-box attacks publicly available, we argue that it is much more beneficial to the defender side than to potential attackers due to the following reasons:
First, since our goal in this work is to derive the most robust classifier, we use a white-box adversary.
While using a white-box adversary for classifier hardening is perfectly fine as the model owner has direct access to it, it is very unusual for a potential attacker to have direct access to the gradients of a deployed model.
Second, together with the attacks, we additionally publish the source code of the novel training scheme, which uses training on adversarial latent space vectors as well as on adversarial discretized domains to significantly improve robustness.
Thereby, model owners can easily harden their models against the published attacks.

As far as data protection is concerned, we do not collect any personally identifiable information in the course of our work.
Data collected from the university and the company network contain only the queried domain names from NX-traffic.
During the monitoring process, we omit all other data so that we only receive a list of the domains that were queried during the recording periods, without reference to the users within the networks.
Furthermore, we observe NX-traffic, which is less privacy-critical compared to resolving DNS traffic, as it usually does not contain user-entered domains, with the exception of typo domains.

\section{Evaluation}
\label{sec:evaluation}
In Fig.~\ref{fig:ATResultsSummaryPlus}, we present our comprehensive results in a heat map that summarizes all evaluation runs of the first fold in a single figure.
It visualizes the FNRs induced by the different attacks on the models trained and hardened using the different methods.

\begin{figure*}[p] 
	\centering
	\resizebox{!}{\textheight}{
		\input{figures/absolute_at_results_fold_0_plus.pgf}
	}
	\caption{FNRs of the classifiers trained using different adversarial training schemes against the attacks we developed. The FNRs against the held-out attacks are framed in blue. None of the classifiers were trained on the black-box DGA samples.}
	\label{fig:ATResultsSummaryPlus}
\end{figure*}

The x-axis specifies the training procedure used to train a classifier, while the y-axis denotes the attacks used.
The first column presents the baseline, i.e., the conventionally trained model without any hardening.
The following three columns contain the results of the models hardened with only discrete attacks, hardened with only attacks from the embedding space, and hardened with our proposed joint AT scheme.
All subsequent columns contain the results of our LOGO evaluation, i.e., a classifier is hardened using joint AT on all attacks except the one depicted on the x-axis, which is excluded from the hardening process.
In the figure, the FNRs against the held-out attacks are highlighted by a blue frame.

We perform the LOGO evaluation to assess the generalization of adversarial robustness across attacks.
When an attack is left out, the number of samples generated by the other attacks automatically increases proportionally.
We perform a LOGO evaluation for each attack group: One for each embedding-space attack, one for each discretization scheme, and one for each discrete domain attack.
We do not perform pairwise LOGO evaluations (e.g., only leave out PGD $L_2$ with LBF $D_c$ discretization) as we want to test the impact of including a new discretization scheme or attack, not just the inclusion of a new pairing.

\subsection{Baseline Robustness}
We briefly discuss the effectiveness of the attacks on the unhardened model.
To analyze its robustness, we examine the FNRs induced by the various attacks on the baseline (first column of Fig.~\ref{fig:ATResultsSummaryPlus}).

\subsubsection{Embedding-Space Attacks}
We group the attacks acting on the embedding space by the different discretization algorithms under the different distance metrics.
The first row of each group shows the vulnerability of the classifiers against adversarial embedding vectors.
All embedding space attacks are very effective and result in FNRs between 98\% and 100\%.
However, these adversarial embedding vectors are of little use to an adversary as they are likely never reachable in practice.

\subsubsection{Discretized Embedding-Space Attacks}
\label{sec:desa}
The discretization to bridge the gap between embedding space vectors and valid domains seems promising:
For every embedding-space attack except C\&W $L_2$, there is at least one discretization scheme with an induced FNR of 100\% (the best scheme for C\&W $L_2$ achieves a FNR of 98\%).

In Appendix~\ref{sec:appendix_discretized_embedding_space_attacks}, we additionally present the success rates of the attacks as a function of the attack strengths and discretization schemes.
Apart from the FNR, we also recorded additional metrics such as the percentage of unique domains, the distances to the original domains in the embedding space, the Levenshtein distances, the model's confidence in the adversarial domains, and the percentage of useable adversarial domains.
We provide an excerpt of these metrics in Table~\ref{tab:EmbeddingSpaceRobustnessAttackKPIs} in the appendix.

It is also prominent that LCO-based rounding is unsuitable in combination with the C\&W $L_2$ attack.
This can be explained by the fact that the C\&W attack is an optimization-based technique that tries to find the smallest perturbation w.r.t. the $L_2$ norm that fools the classifier to a certain degree of confidence.
This is different from the other attacks we examine, such as PGD, which does not have a penalty term for the distance from the origin.
This results in the LCO attacks almost not inducing any changes:
Even at the highest tested confidence setting, the mean Levenshtein distance between input domains and adversarial domains is at most $0.63$.
LBF attacks do not have that same limitation, as they can induce changes simply by varying the domain length.

Moreover, it can be seen that $D_c$-based rounding works best for LCO attacks achieving highest success rates. 
On the other hand, $L_\infty$-based rounding should be avoided.

\subsubsection{Discrete Attacks}
The discrete attacks are also very effective and reach FNRs above 99\%.
HotFlip is able to induce a FNR of $94.04 \%$ by just flipping one character and $99.48 \%$ with two flips, increasing to $99.80 \%$ at five flips.
MaskDGA-WB also increases the FNR of the classifier to over $99\%$.
Similar to the discretized embedding-space attacks we present additional statistics for the discrete attacks in the Appendix~\ref{sec:appendix_discrete_attacks}.

\subsection{Classifiers Hardening}
\label{sec:evaluation_defend}
Columns two, three, and four of Fig.\ref{fig:ATResultsSummaryPlus} summarize the results for the differently hardened classifiers.

\subsubsection{Training on Adversarial Domains}
Training on adversarial domains increases robustness across almost all discrete attacks compared to the unhardened classifier with varying degrees of success.
The robustness against LCO attacks increases significantly.
Such attacks only reach FNRs of at most $69\%$ on the hardened classifier.
The same does, however, not hold for other attacks:
Robustness against the LBF attacks only increases marginally, with PGD $L_\infty$ LBF $L_2$ still reaching a FNR of 100\%.
An intriguing exception are C\&W $L_2$ attacks, whose FNRs are reduced to less than $32\%$ across all discretization schemes. 
Robustness against embedding-space attacks is unaffected.
However, this is not critical, as attackers can never directly use adversarial embedding vectors.

\subsubsection{Training on Adversarial Embedding Vectors}
Training on adversarial embedding vectors increases the robustness against embedding space attacks and against all discretized embedding-space attacks except for C\&W $L_2$.
Interestingly, the classifier hardened in this way frequently achieves a higher robustness against discrete domain attacks than the classifier that is directly trained on these attacks.
Nevertheless, we also observe the opposite for C\&W $L_2$ attacks.
Additionally, training on adversarial embedding vectors helps only slightly against MaskDGA-WB.

\subsubsection{Joint Adversarial Training}
Our results indicate that training on adversarial embedding vectors and adversarial domains together is key to successfully hardening a DGA classifier:
The classifier hardened in this way is more robust against the majority of discrete attacks than both the classifier hardened only against adversarial domains and that only hardened against adversarial embedding vectors.
Across all discrete white-box attacks, the jointly hardened classifier achieves, on average, a $10.15\%$ better FNR than the classifier trained on embedding-space attacks, and a $3.6\%$ better FNR than the classifier trained on discrete domains.

Nevertheless, HotFlip still manages to outwit the classifier reaching a FNR of 96\%.
Moreover, both BAT variants and PGD $L_\infty$ paired with LBF $D_c$ discretization reach FNRs between 94\% and 95\%.
The jointly trained classifier also does not seem to be very robust against embedding-space attacks, which is not critical in practice.

\subsection{Leave One Group Out Evaluation}
We perform a LOGO evaluation to analyze how well the classifier generalizes to unknown attacks.
To this end, we focus on robustness relative to the classifier hardened using joint AT on all attacks.

Leaving out PGD $L_2$ leads to decreased robustness against the PGD and BAT-based attacks.
The robustness against C\&W $L_2$, HotFlip, and MaskDGA-WB remains almost unaffected.
Training without PGD $L_\infty$ has only a minor influence on the robustness against the same attack.
Moreover, leaving it out seemingly increases the robustness against adversarial embedding vectors and BAT-based attacks significantly.
Omitting one of the BAT-based attacks reduces the robustness the most for PGD $L_2$, for the other attacks only small differences are measurable.
Leaving out C\&W $L_2$ does not significantly affect robustness against the C\&W $L_2$ attack but slightly reduces the robustness against PGD and BAT-based attacks.

With respect to the discretization schemes, omitting LBF $L_2$ or LCO $D_c$ reduces robustness to a lesser extent than omitting any of the other three schemes over a wide range of attacks.
Hence, we find that there are non-insignificant interactions between the discretization schemes.
Further, it is striking that omitting LCO $D_c$ discretization drastically reduces the robustness against C\&W~$L_2$~$D_c$.

Looking at the NLP attacks, leaving out MaskDGA-WB reduces robustness against MaskDGA-WB.
It, however, also meaningfully reduces accuracy against the pre-computed black-box samples provided by the authors.
Leaving out HotFlip increases the FNR against HotFlip by 3\%, which is significant as HotFlip reaches a FNR of 96\% against the jointly hardened classifier.
For the black-box attacks, the FNRs induced by MaskDGA-BB vary greatly, while the other attacks are only slightly affected by the omission of an attack.

\subsection{Investigating Anomalies}
\label{sec:investigating_anomalies}
Some anomalies occurred in the previous section, which we examine in more detail here.
To investigate these anomalies, we take a closer look on the results of all five folds.
For the sake of completeness, we present the FNRs for the baseline and for the model hardened by joint AT across all five folds in Fig.~\ref{fig:ATCrossFoldPlus} in the appendix.
Additionally, for the anomalies we encountered during the LOGO evaluation, we present the FNRs when we omit a particular attack.

When analyzing the results of the first fold, we noticed that training without PGD $L_\infty$ significantly increases robustness against the BAT-based attacks.
If we analyze the results across all folds, we only observe this behavior in the first two folds.
For the other three folds, the jointly hardened classifier is similarly robust compared to the scenario in which PGD $L_\infty$ is omitted.
In addition, in these three folds the jointly hardened classifier is also remarkably robust against adversarial vectors.
We reckon that this behavior is due to the fact that these folds ended in two different local optima.

Additionally, we also observed a large robustness decrease against C\&W $L_2$ with LBF $D_c$ discretization when leaving out LBF $D_c$ discretization during the first fold.
Here, the other four folds do not exhibit robustness decreases of the same magnitude.

Hence, the take away message is that when hardening DGA classifiers, it is important to re-evaluate the hardened classifier to ensure that the model has not converged to a local optimum that is still vulnerable.
We reckon that weighting the two different types of minibatches in joint AT could lead to more consistent results, as the increased robustness against the BAT-based attacks is accompanied by the increased robustness against adversarial vectors.

\subsection{Analysis of Generated Adversarial Domains}
\label{sec:evaluation_bias}
In~\cite{drichel_false_2023}, explainability techniques were used to identify several biases of existing DGA classifiers.
In this section, we experiment with a novel way of identifying loopholes:
Analyzing the properties of the generated adversarial domains might reveal systematic weaknesses of the DGA classifiers that can be abused outside of costly white-box attacks.
We identify two loopholes and develop simple attacks that misuse them to effectively circumvent the classifier.
To this end, we analyze the character and length distribution of the benign, malicious, and successful AEs generated on the unhardened classifier.

A manual dataset investigation reveals that the hyphen often appears in adversarial domains.
We see that the university dataset contains a decent amount of hyphens, but in DGArchive there are almost no hyphens.
Our adversarial attacks seem to have found this weakness and exploited it to the fullest:
The hyphen is the most popular character for AEs.
To validate this theory, we devise a new attack algorithm:
\emph{HyphenDGA} is a black-box attack that randomly replaces roughly half of all characters with hyphens, while taking care not to violate the rules of fully valid e2LDs.
The attack is astonishingly successful:
It reaches an average FNR of 99.9\% across all folds, which puts it on par with the gradient-based white-box attacks we developed.
We perform a similar analysis for the frequency of n-grams and additionally for the frequency of SentencePiece~\cite{Kudo2018} tokens but do not find any additional strong~outliers.

Nevertheless, one interesting observation is the frequent occurrence of the character ``i'' in our AEs.
We find that ``i'' is the closest character to the padding character in the $L_2$ norm with respect to the embedding matrix across three of our five folds.
This means that especially $L_2$ LBF attacks tend to round characters that go beyond the original domain length to an ``i'' if the underlying attack made no significant changes to that position.
We additionally notice, that some of our attacks generated very long domain names.
On the other hand, there are nearly no domains in DGArchive and the university data whose e2LD has more than 35 characters.
However, many adversarial domains generated by the LBF discretization approaches have lengths above 40 characters.
To validate that this bias can be easily exploited, we develop \emph{LengthDGA}.
LengthDGA extends a domain to $48$ characters by prepending the character ``i'' as often as required.
We choose the target length of $48$, as it is the most frequent length among successful AEs following the minimum required length of seven.
Additionally, we choose the character ``i'' as that is the second most popular character after the hyphen.
LengthDGA reaches an average induced FNR of 96.0\%.

Now we evaluate these DGAs against the jointly hardened classifiers across all five folds to assess the resilience of the hardened classifiers to the revealed biases.
The hardened model performs much better than the unhardened model, although we did not explicitly train with these DGAs.
HyphenDGA leads to a FNR of 40.0\% while LengthDGA only achieves a FNR of 2.4\% (which corresponds to a reduction of 59.9\% and 93.6\% respectively compared to the unhardened model).
This demonstrates that AT is a viable technique to mitigate potential biases in the training data. 

Finally, we go through the same process we used to find the biases to check if AT introduces new biases.
However, we do not find any evident patterns in the generated adversarial domains.
Looking at the character and length distributions of the successful adversarial domain names shows that they have become similar to the distribution of actual benign domains.
\section{Real-World Study}
\label{sec:evaluation_rw}
The real-world performance of the hardened classifiers is of the highest interest.
On the one hand, several works have identified a performance-robustness trade-off.
On the other hand, especially in the realm of NLP, AT has been found to be an effective regularizer and boost the real-world performance.
In this section, we evaluate the real-world performance of the hardened classifier and compare it to the baseline.
In this process, we also evaluate the models' ability to generalize to unseen DGAs and different networks.
Additionally, we investigate whether the classifiers are time-robust. 

\subsection{Real-World Evaluation}
\begin{figure} 
 	\resizebox{\columnwidth}{!}{
		\centering
		\includegraphics[width=1.0\linewidth]{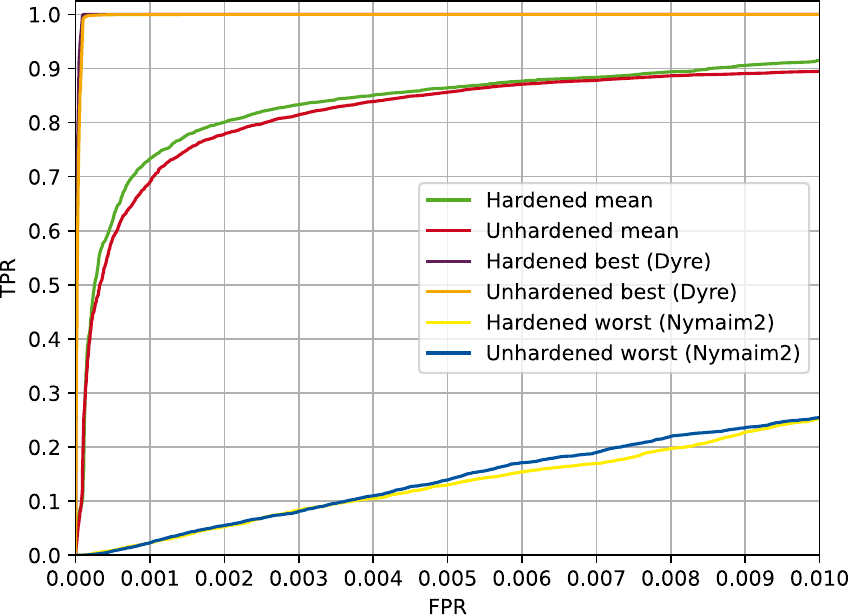}
	}  
	\caption{Averaged ROC curves across five folds of the real-world study. The mean ROC curves are additionally averaged over 46 evaluation runs.}
	\label{fig:RealWorldStudy}
\end{figure}

Recall that all our classifiers are trained with benign e2LDs from the university network until mid-November 2017 and AGDs generated by DGAs before the end of the benign data recording period.
Now, we evaluate all classifiers on the 373 million benign NXDs of the company network from April 2019.
The malicious part of our test set consists of all domains from DGArchive that were generated by DGAs in April 2019.
This includes approximately 1.2 million domains across 46 DGAs.
As it is unclear how many DGAs are present in a real-world network, we try to estimate the worst-case classification performance.
To this end, we perform one evaluation run for each DGA, including all 373 million benign samples and all malicious samples of that particular DGA.
Fig.~\ref{fig:RealWorldStudy} shows the average ROC curve across all five folds and 46 evaluation runs.
In addition, the visualization also shows the ROC curve of the worst and best detected DGA by the classifiers.
We only show the ROC curve until a FPR of $0.01$, as most real-world deployments would target a FPR of less than $1 \%$.

The ROC curve shows that, on average, the hardened classifiers measurably outperform the unhardened classifiers.
To quantify this, we calculate the normalized area under the bounded curves.
The bounded ROC AUC of the unhardened classifier is $0.81335$, and that of the hardened classifier is $0.82894$, which is a $1.56\%$ improvement.
Nevertheless, the ROC curve for the \textit{Nymaim2} DGA shows that this improvement does not apply across all DGAs.
The increased real-world performance could be caused by the strong regularization effect that AT is known for.
All in all, these results show that both classifiers generalize well between different networks and are remarkably time-robust (there is a difference of about 17 months between the samples of the training set and the test set).

\subsection{Capability of Detecting Unknown DGAs}
\label{sec:evaluation_rw_unknown}
Since the time of classifier training $27$ new DGAs have been introduced to DGArchive.
To complete our performance comparison between the hardened and unhardened classifiers, we evaluate the ability of the classifiers to detect unknown DGAs.

On average, the adversarially trained classifiers are $4.7\%$ better at detecting unknown DGAs than the unhardened classifiers.
For 21 out of 27 DGAs the hardened classifiers achieve higher detection scores.
Only for six DGAs the unhardened classifier is on par or better than the hardened classifier.
For the sake of completeness, we also include the average TPRs of the classifiers across all five folds per DGA in Table~\ref{tbl:NovelDGAResults} in the appendix.
In conclusion, our results show that the hardened classifier is better at detecting unknown~DGAs.


\section{Impact \& Discussion}
\label{sec:discussion}
Our aim in this work is to derive classifiers that are as robust as possible for practical use.
It should be noted that context-less approaches are not intended for decision making based on the classification of a single domain, but rather should be integrated into detection systems that enable decision making based on multiple classifications to further reduce FPRs in practice.
The results of our real-world study show that both the unhardened and hardened classifiers are able to detect each DGA to a sufficient degree, even at extremely low FPRs.
The worst detected DGA \textit{Nymaim2} is detected in 10\% of all cases at a FPR of 0.0037 and in 20\% of all cases at a FPR of 0.0082.
On average, the unhardened classifier achieves a TPR of 69.0\% at a FPR of 0.001 and a TPR of 89.5\% at a FPR of 0.01.
In contrast, the hardened classifier achieves TPRs between 73.3\% and 91.5\% at those fixed FPRs.
In a real-world experiment that focused on reducing experimental bias~\cite{pendlebury_tesseract_2019}, we observed that most DGAs (37 out of 46) are detected with TPRs over 90\% at a FPR of 0.0073.
In addition, the classifiers were also able to detect unknown DGAs for which no training samples were included during training, on average with TPRs between 75.8\% and 80.5\%.
Since most DGAs query a large amount of AGDs before finding a registered domain, we argue that all known and unknown DGAs (with the exception of \textit{Chaes}, see Table~\ref{tbl:NovelDGAResults}) would be recognized by a detection system that bases its decision on the results of multiple classification.

In~\cite{drichel_false_2023} it was shown that the bias-reduced classifier performs significantly worse than a FQDN classifier in detecting some specific DGAs.
For these DGAs, the classification depends heavily on the information outside the e2LD.
To solve this problem, a detection system can be used that integrates both a bias-reduced and a FQDN classifier, thereby maintaining the classification performance of the state of the art without being affected by the identified biases.
In fact, a similarly trained FQDN classifier detects \textit{Chaes} with a TPR of 24.9\%.
Hence, combining the hardened classifier with a FQDN classifier similar to the detection system proposed in~\cite{drichel_false_2023} would also allow for the detection of \textit{Chaes}.
Note that the use of the proposed classification system based on ensemble classification does not necessarily increase robustness against adversarial attacks.
By adapting our attacks to target an encompassing surrogate model that includes both a bias-reduced and a FQDN classifier, we are able to reliably bypass the ensemble.

In conclusion, our real-world study has shown that both the unhardened and hardened bias-reduced classifiers generalize well to unknown networks, are remarkably time-robust, and can detect most never-before-seen DGAs with high confidence.
In a direct comparison, the mean TPR of the hardened classifier is always higher than the mean TPR of the unhardened classifier, which can be attributed to the regularization effects of AT.
Moreover, the hardened classifier is, on average, 4.7\% better at detecting unknown~DGAs.

In terms of robustness to adversarial attacks, we have seen that our proposed joint AT is the most beneficial approach for deriving the most robust classifier possible.
We argue that the hardened classifier is remarkably robust against most of the studied attacks that otherwise induce FNRs of $\approx$ 100\% on the unhardened classifier.
Similar to the detection of non-adversarial DGAs, a classification system that includes a hardened classifier and bases its decision on multiple classifications would be able to detect most attacks.
In our comprehensive LOGO study, we found that the majority of left-out attacks are sufficiently detected.
Here, BAT-based attacks as well as C\&W $L_2$ with LBF $D_c$ discretization achieved critical FNRs during our analysis of the first fold.
However, this is due to the fact that the model converged to a susceptible local optimum.
Therefore, we argue that when hardening classifiers, it is of utmost importance to re-evaluate the classifier against the used adversarial attacks to ensure that this does not happen.
The situation is different for PGD $L_\infty$ with $D_c$ discretization and HotFlip.
If they are omitted from the training set, the FNR induced by the attacks is between 95\% and 99\%.
However, it is important to note that in the considered threat model, we use a white-box adversary who has direct access to all model gradients, can adapt attacks, and directly craft AEs on the target model.
In this threat model, unbounded white-box attacks will eventually produce benign samples, assuming the model has enough capacity to represent the true benign data distribution.
Samples that follow the true benign data distribution should of course be classified as benign by the classifier, which puts the high FNRs into perspective. 
Therefore, the threat from these attacks is much lower in practice.

Finally, we identified two biases in the bias-reduced classifiers and showed that AT significantly increases robustness against attacks that exploit these biases, without the need to explicitly train against these attacks.
All in all, our results show that AT is a convincing improvement for bias-reduced DGA classifiers and a necessity for the use of DGA classifiers in practice.

\section{Conclusion}
\label{sec:conclusion}
In this work, we systematically performed a critical analysis of the robustness of state-of-the-art DGA classifiers.
To this end, we implemented 32 white-box attacks, 19 of which are very effective and induce a FNR of $\approx$ 100\% on unhardened classifiers.
In this context, we proposed controllable discretization algorithms that bridge the gap between adversarial vector representation and valid adversarial domains, thereby making a large body of adversarial ML research directly applicable to DGA classification.
We conducted a comprehensive LOGO evaluation which quantified the classifier's robustness against unknown adversarial attacks.
Thereby, we evaluated different AT approaches and proposed a novel training scheme that leverages both adversarial latent space vectors and discretized adversarial domains to significantly improve robustness.
In our evaluation, we emphasized the importance of re-evaluating a hardened classifier to ensure that it converged to a robust local optimum, and we uncovered training biases in state-of-the-art classifiers that can be easily exploited by attackers to bypass detection, but which can be mitigated by AT.
In a real-world evaluation, we demonstrated that DGA classifiers generalize well to unknown networks and are time-robust.
During our study, we did not observe any trade-off between robustness and performance.
On the contrary, AT acts as a kind of regularization that improves the classifier's detection performance for known and unknown DGAs.
To summarize, AT is able to significantly increase robustness against the studied attacks that otherwise induce FNRs of $\approx$ 100\% on unhardened classifiers.

\section*{Availability}
We implemented all the attacks, discretization algorithms, and defenses discussed in this work as a standalone library, which we make publicly available~\footref{sc} to facilitate hardening of DGA classifiers.

\begin{acks}
We thank the Cyber Analysis \& Defense department of Fraunhofer FKIE, Siemens AG, and the IT Center of RWTH Aachen University for providing data for our research.
\end{acks}

\bibliographystyle{ACM-Reference-Format}
\bibliography{bibliography}


\begin{thebibliography}{79}


\ifx \showCODEN    \undefined \def \showCODEN     #1{\unskip}     \fi
\ifx \showDOI      \undefined \def \showDOI       #1{#1}\fi
\ifx \showISBNx    \undefined \def \showISBNx     #1{\unskip}     \fi
\ifx \showISBNxiii \undefined \def \showISBNxiii  #1{\unskip}     \fi
\ifx \showISSN     \undefined \def \showISSN      #1{\unskip}     \fi
\ifx \showLCCN     \undefined \def \showLCCN      #1{\unskip}     \fi
\ifx \shownote     \undefined \def \shownote      #1{#1}          \fi
\ifx \showarticletitle \undefined \def \showarticletitle #1{#1}   \fi
\ifx \showURL      \undefined \def \showURL       {\relax}        \fi
\providecommand\bibfield[2]{#2}
\providecommand\bibinfo[2]{#2}
\providecommand\natexlab[1]{#1}
\providecommand\showeprint[2][]{arXiv:#2}

\bibitem[Akhtar and Mian(2018)]%
        {Akhtar2018}
\bibfield{author}{\bibinfo{person}{Naveed Akhtar} {and} \bibinfo{person}{Ajmal
  Mian}.} \bibinfo{year}{2018}\natexlab{}.
\newblock \showarticletitle{Threat of Adversarial Attacks on Deep Learning in
  Computer Vision: A Survey}.
\newblock \bibinfo{journal}{\emph{IEEE Access}}  \bibinfo{volume}{6}
  (\bibinfo{year}{2018}).
\newblock
\urldef\tempurl%
\url{https://doi.org/10.1109/ACCESS.2018.2807385}
\showDOI{\tempurl}


\bibitem[Akhtar et~al\mbox{.}(2021)]%
        {Akhtar2021}
\bibfield{author}{\bibinfo{person}{Naveed Akhtar}, \bibinfo{person}{Ajmal
  Mian}, \bibinfo{person}{Navid Kardan}, {and} \bibinfo{person}{Mubarak Shah}.}
  \bibinfo{year}{2021}\natexlab{}.
\newblock \showarticletitle{Advances in Adversarial Attacks and Defenses in
  Computer Vision: A Survey}.
\newblock \bibinfo{journal}{\emph{IEEE Access}}  \bibinfo{volume}{9}
  (\bibinfo{year}{2021}).
\newblock
\urldef\tempurl%
\url{https://doi.org/10.1109/ACCESS.2021.3127960}
\showDOI{\tempurl}


\bibitem[Anderson et~al\mbox{.}(2016)]%
        {Anderson2016}
\bibfield{author}{\bibinfo{person}{Hyrum~S. Anderson},
  \bibinfo{person}{Jonathan Woodbridge}, {and} \bibinfo{person}{Bobby Filar}.}
  \bibinfo{year}{2016}\natexlab{}.
\newblock \showarticletitle{DeepDGA: Adversarially-Tuned Domain Generation and
  Detection}. In \bibinfo{booktitle}{\emph{Workshop on Artificial Intelligence
  and Security}}. \bibinfo{publisher}{ACM}.
\newblock
\showISBNx{9781450345736}
\urldef\tempurl%
\url{https://doi.org/10.1145/2996758.2996767}
\showDOI{\tempurl}


\bibitem[Antonakakis et~al\mbox{.}(2010)]%
        {antonakakis_building_2010}
\bibfield{author}{\bibinfo{person}{Manos Antonakakis}, \bibinfo{person}{Roberto
  Perdisci}, \bibinfo{person}{David Dagon}, \bibinfo{person}{Wenke Lee}, {and}
  \bibinfo{person}{Nick Feamster}.} \bibinfo{year}{2010}\natexlab{}.
\newblock \showarticletitle{{Building a Dynamic Reputation System for DNS}}. In
  \bibinfo{booktitle}{\emph{USENIX Security Symposium}}.
  \bibinfo{publisher}{USENIX Association}.
\newblock
\newblock
\shownote{\url{https://www.usenix.org/legacy/events/sec10/tech/full_papers/Antonakakis.pdf}}.


\bibitem[Antonakakis et~al\mbox{.}(2011)]%
        {antonakakis_detecting_2011}
\bibfield{author}{\bibinfo{person}{Manos Antonakakis}, \bibinfo{person}{Roberto
  Perdisci}, \bibinfo{person}{Wenke Lee}, \bibinfo{person}{Nikolaos~Vasiloglou
  II}, {and} \bibinfo{person}{David Dagon}.} \bibinfo{year}{2011}\natexlab{}.
\newblock \showarticletitle{{Detecting Malware Domains at the Upper DNS
  Hierarchy}}. In \bibinfo{booktitle}{\emph{USENIX Security Symposium}}.
  \bibinfo{publisher}{USENIX Association}.
\newblock
\urldef\tempurl%
\url{https://www.usenix.org/conference/usenix-security-11/detecting-malware-domains-upper-dns-hierarchy}
\showURL{%
\tempurl}


\bibitem[Antonakakis et~al\mbox{.}(2012)]%
        {antonakakis_throwaway_2012}
\bibfield{author}{\bibinfo{person}{Manos Antonakakis}, \bibinfo{person}{Roberto
  Perdisci}, \bibinfo{person}{Yacin Nadji}, \bibinfo{person}{Nikolaos
  Vasiloglou}, \bibinfo{person}{Saeed {Abu-Nimeh}}, \bibinfo{person}{Wenke
  Lee}, {and} \bibinfo{person}{David Dagon}.} \bibinfo{year}{2012}\natexlab{}.
\newblock \showarticletitle{{From Throw-Away Traffic to Bots: Detecting the
  Rise of {DGA}-Based Malware}}. In \bibinfo{booktitle}{\emph{{{USENIX Security
  Symposium}}}}. \bibinfo{publisher}{USENIX Association}.
\newblock
\newblock
\shownote{\url{https://www.usenix.org/conference/usenixsecurity12/technical-sessions/presentation/antonakakis}}.


\bibitem[Athalye et~al\mbox{.}(2018)]%
        {Athalye2018}
\bibfield{author}{\bibinfo{person}{Anish Athalye}, \bibinfo{person}{Nicholas
  Carlini}, {and} \bibinfo{person}{David Wagner}.}
  \bibinfo{year}{2018}\natexlab{}.
\newblock \showarticletitle{Obfuscated Gradients Give a False Sense of
  Security: Circumventing Defenses to Adversarial Examples}. In
  \bibinfo{booktitle}{\emph{International Conference on Machine Learning}},
  \bibfield{editor}{\bibinfo{person}{Jennifer Dy} {and}
  \bibinfo{person}{Andreas Krause}} (Eds.). \bibinfo{publisher}{PMLR}.
\newblock
\urldef\tempurl%
\url{https://proceedings.mlr.press/v80/athalye18a.html}
\showURL{%
\tempurl}


\bibitem[Bai et~al\mbox{.}(2021)]%
        {Bai2021}
\bibfield{author}{\bibinfo{person}{Tao Bai}, \bibinfo{person}{Jinqi Luo},
  \bibinfo{person}{Jun Zhao}, \bibinfo{person}{Bihan Wen}, {and}
  \bibinfo{person}{Qian Wang}.} \bibinfo{year}{2021}\natexlab{}.
\newblock \showarticletitle{Recent Advances in Adversarial Training for
  Adversarial Robustness}. In \bibinfo{booktitle}{\emph{International Joint
  Conference on Artificial Intelligence}}. \bibinfo{publisher}{International
  Joint Conferences on Artificial Intelligence Organization}.
\newblock
\urldef\tempurl%
\url{https://doi.org/10.24963/ijcai.2021/591}
\showDOI{\tempurl}


\bibitem[Bilge et~al\mbox{.}(2014)]%
        {bilge_exposure_2014}
\bibfield{author}{\bibinfo{person}{Leyla Bilge}, \bibinfo{person}{Sevil Sen},
  \bibinfo{person}{Davide Balzarotti}, \bibinfo{person}{Engin Kirda}, {and}
  \bibinfo{person}{Christopher Kruegel}.} \bibinfo{year}{2014}\natexlab{}.
\newblock \showarticletitle{{Exposure: A Passive DNS Analysis Service to Detect
  and Report Malicious Domains}}.
\newblock \bibinfo{journal}{\emph{Transactions on Information and System
  Security}} \bibinfo{volume}{16}, \bibinfo{number}{4}, Article
  \bibinfo{articleno}{14} (\bibinfo{year}{2014}).
\newblock
\newblock
\shownote{\url{https://doi.org/10.1145/2584679}}.


\bibitem[Carlini(2019)]%
        {Carlini2019a}
\bibfield{author}{\bibinfo{person}{Nicholas Carlini}.}
  \bibinfo{year}{2019}\natexlab{}.
\newblock \bibinfo{title}{A Complete List of All (arXiv) Adversarial Example
  Papers}.
\newblock
\newblock
\urldef\tempurl%
\url{https://nicholas.carlini.com/writing/2019/all-adversarial-example-papers.html}
\showURL{%
\tempurl}
\newblock
\shownote{online, accessed 2023-11-28}.


\bibitem[Carlini et~al\mbox{.}(2019)]%
        {Carlini2019}
\bibfield{author}{\bibinfo{person}{Nicholas Carlini}, \bibinfo{person}{Anish
  Athalye}, \bibinfo{person}{Nicolas Papernot}, \bibinfo{person}{Wieland
  Brendel}, \bibinfo{person}{Jonas Rauber}, \bibinfo{person}{Dimitris Tsipras},
  \bibinfo{person}{Ian Goodfellow}, \bibinfo{person}{Aleksander Madry}, {and}
  \bibinfo{person}{Alexey Kurakin}.} \bibinfo{year}{2019}\natexlab{}.
\newblock \showarticletitle{On Evaluating Adversarial Robustness}.
  \bibinfo{publisher}{arXiv:1902.06705}.
\newblock
\urldef\tempurl%
\url{https://doi.org/10.48550/ARXIV.1902.06705}
\showDOI{\tempurl}


\bibitem[Carlini and Wagner(2017a)]%
        {Carlini2017}
\bibfield{author}{\bibinfo{person}{Nicholas Carlini} {and}
  \bibinfo{person}{David Wagner}.} \bibinfo{year}{2017}\natexlab{a}.
\newblock \showarticletitle{Adversarial Examples Are Not Easily Detected:
  Bypassing Ten Detection Methods}. In \bibinfo{booktitle}{\emph{Workshop on
  Artificial Intelligence and Security}}. \bibinfo{publisher}{ACM}.
\newblock
\showISBNx{9781450352024}
\urldef\tempurl%
\url{https://doi.org/10.1145/3128572.3140444}
\showDOI{\tempurl}


\bibitem[Carlini and Wagner(2017b)]%
        {Carlini2017a}
\bibfield{author}{\bibinfo{person}{Nicholas Carlini} {and}
  \bibinfo{person}{David Wagner}.} \bibinfo{year}{2017}\natexlab{b}.
\newblock \showarticletitle{Towards Evaluating the Robustness of Neural
  Networks}. In \bibinfo{booktitle}{\emph{Symposium on Security and Privacy}}.
  \bibinfo{publisher}{IEEE}.
\newblock
\showISSN{2375-1207}
\urldef\tempurl%
\url{https://doi.org/10.1109/SP.2017.49}
\showDOI{\tempurl}


\bibitem[Corley et~al\mbox{.}(2019)]%
        {Corley2019}
\bibfield{author}{\bibinfo{person}{Isaac Corley}, \bibinfo{person}{Jonathan
  Lwowski}, {and} \bibinfo{person}{Justin Hoffman}.}
  \bibinfo{year}{2019}\natexlab{}.
\newblock \showarticletitle{DomainGAN: Generating Adversarial Examples to
  Attack Domain Generation Algorithm Classifiers}.
  \bibinfo{publisher}{arXiv:1911.06285}.
\newblock
\urldef\tempurl%
\url{https://doi.org/10.48550/arxiv.1911.06285}
\showDOI{\tempurl}


\bibitem[Croce and Hein(2020a)]%
        {Croce2020a}
\bibfield{author}{\bibinfo{person}{Francesco Croce} {and}
  \bibinfo{person}{Matthias Hein}.} \bibinfo{year}{2020}\natexlab{a}.
\newblock \showarticletitle{Provable robustness against all adversarial
  $l_p$-perturbations for $p\geq 1$}. In
  \bibinfo{booktitle}{\emph{International Conference on Learning
  Representations}}. \bibinfo{publisher}{OpenReview.net}.
\newblock
\urldef\tempurl%
\url{https://openreview.net/forum?id=rklk\_ySYPB}
\showURL{%
\tempurl}


\bibitem[Croce and Hein(2020b)]%
        {Croce2020}
\bibfield{author}{\bibinfo{person}{Francesco Croce} {and}
  \bibinfo{person}{Matthias Hein}.} \bibinfo{year}{2020}\natexlab{b}.
\newblock \showarticletitle{Reliable Evaluation of Adversarial Robustness with
  an Ensemble of Diverse Parameter-free Attacks}. In
  \bibinfo{booktitle}{\emph{International Conference on Machine Learning}}.
  \bibinfo{publisher}{PMLR}.
\newblock
\urldef\tempurl%
\url{https://proceedings.mlr.press/v119/croce20b.html}
\showURL{%
\tempurl}


\bibitem[Drichel et~al\mbox{.}(2021)]%
        {drichel_first_2021}
\bibfield{author}{\bibinfo{person}{Arthur Drichel}, \bibinfo{person}{Nils
  Faerber}, {and} \bibinfo{person}{Ulrike Meyer}.}
  \bibinfo{year}{2021}\natexlab{}.
\newblock \showarticletitle{{First Step Towards EXPLAINable DGA Multiclass
  Classification}}. In \bibinfo{booktitle}{\emph{International Conference on
  Availability, Reliability and Security}}. \bibinfo{publisher}{ACM}.
\newblock
\newblock
\shownote{\url{https://doi.org/10.1145/3465481.3465749}}.


\bibitem[Drichel and Meyer(2023)]%
        {drichel_false_2023}
\bibfield{author}{\bibinfo{person}{Arthur Drichel} {and}
  \bibinfo{person}{Ulrike Meyer}.} \bibinfo{year}{2023}\natexlab{}.
\newblock \showarticletitle{{False Sense of Security: Leveraging XAI to Analyze
  the Reasoning and True Performance of Context-less DGA Classifiers}}. In
  \bibinfo{booktitle}{\emph{International Symposium on Research in Attacks,
  Intrusions and Defenses}}. \bibinfo{publisher}{ACM}.
\newblock
\newblock
\shownote{\url{https://doi.org/10.1145/3607199.3607231}}.


\bibitem[Drichel et~al\mbox{.}(2020a)]%
        {drichel_analyzing_2020}
\bibfield{author}{\bibinfo{person}{Arthur Drichel}, \bibinfo{person}{Ulrike
  Meyer}, \bibinfo{person}{Samuel Sch\"uppen}, {and} \bibinfo{person}{Dominik
  Teubert}.} \bibinfo{year}{2020}\natexlab{a}.
\newblock \showarticletitle{{Analyzing the Real-World Applicability of {DGA}
  Classifiers}}. In \bibinfo{booktitle}{\emph{International Conference on
  Availability, Reliability and Security}}. \bibinfo{publisher}{ACM}.
\newblock
\newblock
\shownote{\url{https://doi.org/10.1145/3407023.3407030}}.


\bibitem[Drichel et~al\mbox{.}(2020b)]%
        {drichel_making_2020}
\bibfield{author}{\bibinfo{person}{Arthur Drichel}, \bibinfo{person}{Ulrike
  Meyer}, \bibinfo{person}{Samuel Sch\"uppen}, {and} \bibinfo{person}{Dominik
  Teubert}.} \bibinfo{year}{2020}\natexlab{b}.
\newblock \showarticletitle{{Making Use of {NXt} to Nothing: Effect of Class
  Imbalances on {DGA} Detection Classifiers}}. In
  \bibinfo{booktitle}{\emph{International Conference on Availability,
  Reliability and Security}}. \bibinfo{publisher}{ACM}.
\newblock
\newblock
\shownote{\url{https://doi.org/10.1145/3407023.3409190}}.


\bibitem[Ebrahimi et~al\mbox{.}(2018)]%
        {Ebrahimi2017}
\bibfield{author}{\bibinfo{person}{Javid Ebrahimi}, \bibinfo{person}{Anyi Rao},
  \bibinfo{person}{Daniel Lowd}, {and} \bibinfo{person}{Dejing Dou}.}
  \bibinfo{year}{2018}\natexlab{}.
\newblock \showarticletitle{{H}ot{F}lip: White-Box Adversarial Examples for
  Text Classification}. In \bibinfo{booktitle}{\emph{Annual Meeting of the
  Association for Computational Linguistics}}. \bibinfo{publisher}{Association
  for Computational Linguistics}.
\newblock
\urldef\tempurl%
\url{https://doi.org/10.18653/v1/P18-2006}
\showDOI{\tempurl}


\bibitem[Goodfellow et~al\mbox{.}(2015)]%
        {Goodfellow2014}
\bibfield{author}{\bibinfo{person}{Ian~J. Goodfellow},
  \bibinfo{person}{Jonathon Shlens}, {and} \bibinfo{person}{Christian
  Szegedy}.} \bibinfo{year}{2015}\natexlab{}.
\newblock \showarticletitle{Explaining and Harnessing Adversarial Examples}. In
  \bibinfo{booktitle}{\emph{International Conference on Learning
  Representations}}.
\newblock
\urldef\tempurl%
\url{https://doi.org/10.48550/ARXIV.1412.6572}
\showDOI{\tempurl}


\bibitem[Gould et~al\mbox{.}(2020)]%
        {Gould2020}
\bibfield{author}{\bibinfo{person}{Nathaniel Gould}, \bibinfo{person}{Taishi
  Nishiyama}, {and} \bibinfo{person}{Kazunori Kamiya}.}
  \bibinfo{year}{2020}\natexlab{}.
\newblock \showarticletitle{Domain Generation Algorithm Detection Utilizing
  Model Hardening Through GAN-Generated Adversarial Examples}. In
  \bibinfo{booktitle}{\emph{Deployable Machine Learning for Security Defense}}.
  \bibinfo{publisher}{Springer}.
\newblock
\urldef\tempurl%
\url{https://doi.org/10.1007/978-3-030-59621-7_5}
\showDOI{\tempurl}


\bibitem[Grill et~al\mbox{.}(2015)]%
        {grill_detecting_2015}
\bibfield{author}{\bibinfo{person}{Martin Grill}, \bibinfo{person}{Ivan
  Nikolaev}, \bibinfo{person}{Veronica Valeros}, {and} \bibinfo{person}{Martin
  Rehak}.} \bibinfo{year}{2015}\natexlab{}.
\newblock \showarticletitle{{Detecting {DGA} Malware Using NetFlow}}. In
  \bibinfo{booktitle}{\emph{IFIP/IEEE Integrated Network Management}}.
  \bibinfo{publisher}{{IEEE}}.
\newblock
\newblock
\shownote{\url{https://doi.org/10.1109/INM.2015.7140486}}.


\bibitem[Guo et~al\mbox{.}(2018)]%
        {Guo2018}
\bibfield{author}{\bibinfo{person}{Chuan Guo}, \bibinfo{person}{Mayank Rana},
  \bibinfo{person}{Moustapha Ciss{\'{e}}}, {and} \bibinfo{person}{Laurens
  van~der Maaten}.} \bibinfo{year}{2018}\natexlab{}.
\newblock \showarticletitle{Countering Adversarial Images using Input
  Transformations}. In \bibinfo{booktitle}{\emph{International Conference on
  Learning Representations}}. \bibinfo{publisher}{OpenReview.net}.
\newblock
\urldef\tempurl%
\url{https://openreview.net/forum?id=SyJ7ClWCb}
\showURL{%
\tempurl}


\bibitem[Hu et~al\mbox{.}(2023)]%
        {Hu2023}
\bibfield{author}{\bibinfo{person}{Xiaoyan Hu}, \bibinfo{person}{Hao Chen},
  \bibinfo{person}{Miao Li}, \bibinfo{person}{Guang Cheng},
  \bibinfo{person}{Ruidong Li}, \bibinfo{person}{Hua Wu}, {and}
  \bibinfo{person}{Yali Yuan}.} \bibinfo{year}{2023}\natexlab{}.
\newblock \showarticletitle{ReplaceDGA: BiLSTM-Based Adversarial DGA With High
  Anti-Detection Ability}.
\newblock \bibinfo{journal}{\emph{Transactions on Information Forensics and
  Security}}  \bibinfo{volume}{18} (\bibinfo{year}{2023}).
\newblock
\urldef\tempurl%
\url{https://doi.org/10.1109/TIFS.2023.3293956}
\showDOI{\tempurl}


\bibitem[Jin et~al\mbox{.}(2020)]%
        {Jin2020}
\bibfield{author}{\bibinfo{person}{Di Jin}, \bibinfo{person}{Zhijing Jin},
  \bibinfo{person}{Joey~Tianyi Zhou}, {and} \bibinfo{person}{Peter Szolovits}.}
  \bibinfo{year}{2020}\natexlab{}.
\newblock \showarticletitle{Is {BERT} Really Robust? {A} Strong Baseline for
  Natural Language Attack on Text Classification and Entailment}. In
  \bibinfo{booktitle}{\emph{{AAAI} Conference on Artificial Intelligence}}.
  \bibinfo{publisher}{{AAAI} Press}.
\newblock
\urldef\tempurl%
\url{https://doi.org/10.1609/aaai.v34i05.6311}
\showDOI{\tempurl}


\bibitem[Kingma and Ba(2015)]%
        {Kingma2017}
\bibfield{author}{\bibinfo{person}{Diederik~P. Kingma} {and}
  \bibinfo{person}{Jimmy Ba}.} \bibinfo{year}{2015}\natexlab{}.
\newblock \showarticletitle{Adam: {A} Method for Stochastic Optimization}. In
  \bibinfo{booktitle}{\emph{International Conference on Learning
  Representations}}. \bibinfo{publisher}{OpenReview.net}.
\newblock
\urldef\tempurl%
\url{https://doi.org/10.48550/ARXIV.1412.6980}
\showDOI{\tempurl}
\showeprint[arxiv]{1412.6980}


\bibitem[Klensin(2010)]%
        {rfc5890}
\bibfield{author}{\bibinfo{person}{Dr. John~C. Klensin}.}
  \bibinfo{year}{2010}\natexlab{}.
\newblock \bibinfo{title}{{Internationalized Domain Names for Applications
  (IDNA): Definitions and Document Framework}}.
\newblock \bibinfo{howpublished}{RFC 5890}.
\newblock
\urldef\tempurl%
\url{https://doi.org/10.17487/RFC5890}
\showDOI{\tempurl}


\bibitem[Kreuk et~al\mbox{.}(2018)]%
        {Kreuk2018}
\bibfield{author}{\bibinfo{person}{Felix Kreuk}, \bibinfo{person}{Assi Barak},
  \bibinfo{person}{Shir Aviv-Reuven}, \bibinfo{person}{Moran Baruch},
  \bibinfo{person}{Benny Pinkas}, {and} \bibinfo{person}{Joseph Keshet}.}
  \bibinfo{year}{2018}\natexlab{}.
\newblock \showarticletitle{Deceiving End-to-End Deep Learning Malware
  Detectors using Adversarial Examples}. \bibinfo{publisher}{arXiv:1802.04528}.
\newblock
\urldef\tempurl%
\url{https://doi.org/10.48550/ARXIV.1802.04528}
\showDOI{\tempurl}


\bibitem[Kudo and Richardson(2018)]%
        {Kudo2018}
\bibfield{author}{\bibinfo{person}{Taku Kudo} {and} \bibinfo{person}{John
  Richardson}.} \bibinfo{year}{2018}\natexlab{}.
\newblock \showarticletitle{{S}entence{P}iece: A simple and language
  independent subword tokenizer and detokenizer for Neural Text Processing}. In
  \bibinfo{booktitle}{\emph{Conference on Empirical Methods in Natural Language
  Processing}}. \bibinfo{publisher}{Association for Computational Linguistics}.
\newblock
\urldef\tempurl%
\url{https://doi.org/10.18653/v1/D18-2012}
\showDOI{\tempurl}


\bibitem[Kurakin et~al\mbox{.}(2017a)]%
        {Kurakin2018}
\bibfield{author}{\bibinfo{person}{Alexey Kurakin}, \bibinfo{person}{Ian~J
  Goodfellow}, {and} \bibinfo{person}{Samy Bengio}.}
  \bibinfo{year}{2017}\natexlab{a}.
\newblock \showarticletitle{Adversarial examples in the physical world}. In
  \bibinfo{booktitle}{\emph{International Conference on Learning
  Representations}}. \bibinfo{publisher}{OpenReview.net}.
\newblock
\urldef\tempurl%
\url{https://openreview.net/forum?id=HJGU3Rodl}
\showURL{%
\tempurl}


\bibitem[Kurakin et~al\mbox{.}(2017b)]%
        {Kurakin2017}
\bibfield{author}{\bibinfo{person}{Alexey Kurakin}, \bibinfo{person}{Ian~J.
  Goodfellow}, {and} \bibinfo{person}{Samy Bengio}.}
  \bibinfo{year}{2017}\natexlab{b}.
\newblock \showarticletitle{Adversarial Machine Learning at Scale}. In
  \bibinfo{booktitle}{\emph{International Conference on Learning
  Representations}}. \bibinfo{publisher}{OpenReview.net}.
\newblock
\urldef\tempurl%
\url{https://openreview.net/forum?id=BJm4T4Kgx}
\showURL{%
\tempurl}


\bibitem[Le~Pochat et~al\mbox{.}(2019)]%
        {lepochat_tranco_2019}
\bibfield{author}{\bibinfo{person}{Victor Le~Pochat}, \bibinfo{person}{Tom
  Van~Goethem}, \bibinfo{person}{Samaneh Tajalizadehkhoob},
  \bibinfo{person}{Maciej Korczynski}, {and} \bibinfo{person}{Wouter Joosen}.}
  \bibinfo{year}{2019}\natexlab{}.
\newblock \showarticletitle{{Tranco: A Research-Oriented Top Sites Ranking
  Hardened Against Manipulation}}. In \bibinfo{booktitle}{\emph{Network and
  Distributed System Security Symposium}}. \bibinfo{publisher}{{Internet
  Society}}.
\newblock
\newblock
\shownote{\url{https://www.ndss-symposium.org/ndss-paper/tranco-a-research-oriented-top-sites-ranking-hardened-against-manipulation/}}.


\bibitem[Li et~al\mbox{.}(2019)]%
        {Li2019}
\bibfield{author}{\bibinfo{person}{Jinfeng Li}, \bibinfo{person}{Shouling Ji},
  \bibinfo{person}{Tianyu Du}, \bibinfo{person}{Bo Li}, {and}
  \bibinfo{person}{Ting Wang}.} \bibinfo{year}{2019}\natexlab{}.
\newblock \showarticletitle{TextBugger: Generating Adversarial Text Against
  Real-world Applications}. In \bibinfo{booktitle}{\emph{Network and
  Distributed System Security Symposium}}. \bibinfo{publisher}{The Internet
  Society}.
\newblock
\urldef\tempurl%
\url{https://www.ndss-symposium.org/ndss-paper/textbugger-generating-adversarial-text-against-real-world-applications/}
\showURL{%
\tempurl}


\bibitem[Li et~al\mbox{.}(2023)]%
        {Li2023}
\bibfield{author}{\bibinfo{person}{Linyi Li}, \bibinfo{person}{Tao Xie}, {and}
  \bibinfo{person}{Bo Li}.} \bibinfo{year}{2023}\natexlab{}.
\newblock \showarticletitle{SoK: Certified Robustness for Deep Neural
  Networks}. In \bibinfo{booktitle}{\emph{Symposium on Security and Privacy}}.
  \bibinfo{publisher}{IEEE}.
\newblock
\urldef\tempurl%
\url{https://doi.org/10.1109/SP46215.2023.10179303}
\showDOI{\tempurl}


\bibitem[Liu et~al\mbox{.}(2022)]%
        {Liu2022}
\bibfield{author}{\bibinfo{person}{Qihe Liu}, \bibinfo{person}{Gao Yu},
  \bibinfo{person}{Yuanyuan Wang}, {and} \bibinfo{person}{Zeng Yi}.}
  \bibinfo{year}{2022}\natexlab{}.
\newblock \showarticletitle{A Novel DGA Domain Adversarial Sample Generation
  Method By Geometric Perturbation}. In \bibinfo{booktitle}{\emph{International
  Conference on Advanced Information Science and System}}.
  \bibinfo{publisher}{ACM}.
\newblock
\showISBNx{9781450385862}
\urldef\tempurl%
\url{https://doi.org/10.1145/3503047.3503080}
\showDOI{\tempurl}


\bibitem[Liu et~al\mbox{.}(2021)]%
        {Liu2021}
\bibfield{author}{\bibinfo{person}{Wanping Liu}, \bibinfo{person}{Zhoulan
  Zhang}, \bibinfo{person}{Cheng Huang}, {and} \bibinfo{person}{Yong Fang}.}
  \bibinfo{year}{2021}\natexlab{}.
\newblock \showarticletitle{CLETer: A Character-level Evasion Technique Against
  Deep Learning DGA Classifiers}.
\newblock \bibinfo{journal}{\emph{Endorsed Transactions on Security and
  Safety}} \bibinfo{volume}{7}, \bibinfo{number}{24} (\bibinfo{year}{2021}).
\newblock
\urldef\tempurl%
\url{https://doi.org/10.4108/eai.18-2-2021.168723}
\showDOI{\tempurl}


\bibitem[Liu et~al\mbox{.}(2017)]%
        {Liu2016}
\bibfield{author}{\bibinfo{person}{Yanpei Liu}, \bibinfo{person}{Xinyun Chen},
  \bibinfo{person}{Chang Liu}, {and} \bibinfo{person}{Dawn Song}.}
  \bibinfo{year}{2017}\natexlab{}.
\newblock \showarticletitle{Delving into Transferable Adversarial Examples and
  Black-box Attacks}. In \bibinfo{booktitle}{\emph{International Conference on
  Learning Representations}}. \bibinfo{publisher}{OpenReview.net}.
\newblock
\urldef\tempurl%
\url{https://openreview.net/forum?id=Sys6GJqxl}
\showURL{%
\tempurl}


\bibitem[Lucas et~al\mbox{.}(2023)]%
        {Lucas2023}
\bibfield{author}{\bibinfo{person}{Keane Lucas}, \bibinfo{person}{Samruddhi
  Pai}, \bibinfo{person}{Weiran Lin}, \bibinfo{person}{Lujo Bauer},
  \bibinfo{person}{Michael~K. Reiter}, {and} \bibinfo{person}{Mahmood Sharif}.}
  \bibinfo{year}{2023}\natexlab{}.
\newblock \showarticletitle{Adversarial Training for {Raw-Binary} Malware
  Classifiers}. In \bibinfo{booktitle}{\emph{USENIX Security Symposium}}.
  \bibinfo{publisher}{USENIX Association}.
\newblock
\showISBNx{978-1-939133-37-3}
\urldef\tempurl%
\url{https://www.usenix.org/conference/usenixsecurity23/presentation/lucas}
\showURL{%
\tempurl}


\bibitem[Madry et~al\mbox{.}(2018)]%
        {Madry2017}
\bibfield{author}{\bibinfo{person}{Aleksander Madry},
  \bibinfo{person}{Aleksandar Makelov}, \bibinfo{person}{Ludwig Schmidt},
  \bibinfo{person}{Dimitris Tsipras}, {and} \bibinfo{person}{Adrian Vladu}.}
  \bibinfo{year}{2018}\natexlab{}.
\newblock \showarticletitle{Towards Deep Learning Models Resistant to
  Adversarial Attacks}. In \bibinfo{booktitle}{\emph{International Conference
  on Learning Representations}}. \bibinfo{publisher}{OpenReview.net}.
\newblock
\urldef\tempurl%
\url{https://openreview.net/forum?id=rJzIBfZAb}
\showURL{%
\tempurl}


\bibitem[Mockapetris(1987)]%
        {rfc1035}
\bibfield{author}{\bibinfo{person}{Paul Mockapetris}.}
  \bibinfo{year}{1987}\natexlab{}.
\newblock \bibinfo{title}{{Domain names - implementation and specification}}.
\newblock \bibinfo{howpublished}{RFC 1035}.
\newblock
\urldef\tempurl%
\url{https://doi.org/10.17487/RFC1035}
\showDOI{\tempurl}


\bibitem[Nie et~al\mbox{.}(2022)]%
        {nie2022pkdga}
\bibfield{author}{\bibinfo{person}{Lihai Nie}, \bibinfo{person}{Xiaoyang Shan},
  \bibinfo{person}{Laiping Zhao}, {and} \bibinfo{person}{Keqiu Li}.}
  \bibinfo{year}{2022}\natexlab{}.
\newblock \showarticletitle{PKDGA: A Partial Knowledge-based Domain Generation
  Algorithm for Botnets}. \bibinfo{publisher}{arXiv:2212.04234}.
\newblock
\urldef\tempurl%
\url{https://doi.org/10.48550/arXiv.2212.04234}
\showDOI{\tempurl}


\bibitem[Papernot et~al\mbox{.}(2016)]%
        {Papernot2016b}
\bibfield{author}{\bibinfo{person}{Nicolas Papernot}, \bibinfo{person}{Patrick
  McDaniel}, {and} \bibinfo{person}{Ian Goodfellow}.}
  \bibinfo{year}{2016}\natexlab{}.
\newblock \showarticletitle{Transferability in Machine Learning: from Phenomena
  to Black-Box Attacks using Adversarial Samples}.
  \bibinfo{publisher}{arXiv:1605.07277}.
\newblock
\urldef\tempurl%
\url{https://doi.org/10.48550/ARXIV.1605.07277}
\showDOI{\tempurl}


\bibitem[Pardue and Desgats(2023)]%
        {Pardue2023}
\bibfield{author}{\bibinfo{person}{Lucas Pardue} {and} \bibinfo{person}{Julien
  Desgats}.} \bibinfo{year}{2023}\natexlab{}.
\newblock \bibinfo{title}{HTTP/2 Rapid Reset: deconstructing the
  record-breaking attack}.
\newblock
\newblock
\urldef\tempurl%
\url{https://blog.cloudflare.com/technical-breakdown-http2-rapid-reset-ddos-attack/}
\showURL{%
\tempurl}
\newblock
\shownote{online, accessed 2023-11-21}.


\bibitem[Peck et~al\mbox{.}(2019)]%
        {Peck2019}
\bibfield{author}{\bibinfo{person}{Jonathan Peck}, \bibinfo{person}{Claire
  Nie}, \bibinfo{person}{Raaghavi Sivaguru}, \bibinfo{person}{Charles Grumer},
  \bibinfo{person}{Femi Olumofin}, \bibinfo{person}{Bin Yu},
  \bibinfo{person}{Anderson Nascimento}, {and} \bibinfo{person}{Martine
  De~Cock}.} \bibinfo{year}{2019}\natexlab{}.
\newblock \showarticletitle{CharBot: A Simple and Effective Method for Evading
  DGA Classifiers}.
\newblock \bibinfo{journal}{\emph{IEEE Access}}  \bibinfo{volume}{7}
  (\bibinfo{year}{2019}).
\newblock
\urldef\tempurl%
\url{https://doi.org/10.1109/ACCESS.2019.2927075}
\showDOI{\tempurl}


\bibitem[Pendlebury et~al\mbox{.}(2019)]%
        {pendlebury_tesseract_2019}
\bibfield{author}{\bibinfo{person}{Feargus Pendlebury}, \bibinfo{person}{Fabio
  Pierazzi}, \bibinfo{person}{Roberto Jordaney}, \bibinfo{person}{Johannes
  Kinder}, {and} \bibinfo{person}{Lorenzo Cavallaro}.}
  \bibinfo{year}{2019}\natexlab{}.
\newblock \showarticletitle{{{TESSERACT}: Eliminating Experimental Bias in
  Malware Classification across Space and Time}}. In
  \bibinfo{booktitle}{\emph{USENIX Security Symposium}}.
  \bibinfo{publisher}{USENIX Association}.
\newblock
\newblock
\shownote{\url{https://www.usenix.org/conference/usenixsecurity19/presentation/pendlebury}}.


\bibitem[Plohmann et~al\mbox{.}(2016)]%
        {Plohmann2016}
\bibfield{author}{\bibinfo{person}{Daniel Plohmann}, \bibinfo{person}{Khaled
  Yakdan}, \bibinfo{person}{Michael Klatt}, \bibinfo{person}{Johannes Bader},
  {and} \bibinfo{person}{Elmar Gerhards-Padilla}.}
  \bibinfo{year}{2016}\natexlab{}.
\newblock \showarticletitle{A Comprehensive Measurement Study of Domain
  Generating Malware}. In \bibinfo{booktitle}{\emph{USENIX Security
  Symposium}}. \bibinfo{publisher}{USENIX Association}.
\newblock
\showISBNx{978-1-931971-32-4}
\urldef\tempurl%
\url{https://www.usenix.org/conference/usenixsecurity16/technical-sessions/presentation/plohmann}
\showURL{%
\tempurl}


\bibitem[Pruthi et~al\mbox{.}(2019)]%
        {Pruthi2019}
\bibfield{author}{\bibinfo{person}{Danish Pruthi}, \bibinfo{person}{Bhuwan
  Dhingra}, {and} \bibinfo{person}{Zachary~C. Lipton}.}
  \bibinfo{year}{2019}\natexlab{}.
\newblock \showarticletitle{Combating Adversarial Misspellings with Robust Word
  Recognition}. In \bibinfo{booktitle}{\emph{Annual Meeting of the Association
  for Computational Linguistics}}. \bibinfo{publisher}{Association for
  Computational Linguistics}.
\newblock
\urldef\tempurl%
\url{https://doi.org/10.18653/v1/P19-1561}
\showDOI{\tempurl}


\bibitem[Ren et~al\mbox{.}(2019)]%
        {Ren2019}
\bibfield{author}{\bibinfo{person}{Shuhuai Ren}, \bibinfo{person}{Yihe Deng},
  \bibinfo{person}{Kun He}, {and} \bibinfo{person}{Wanxiang Che}.}
  \bibinfo{year}{2019}\natexlab{}.
\newblock \showarticletitle{Generating Natural Language Adversarial Examples
  through Probability Weighted Word Saliency}. In
  \bibinfo{booktitle}{\emph{Annual Meeting of the Association for Computational
  Linguistics}}. \bibinfo{publisher}{Association for Computational
  Linguistics}.
\newblock
\urldef\tempurl%
\url{https://doi.org/10.18653/v1/P19-1103}
\showDOI{\tempurl}


\bibitem[Rossow et~al\mbox{.}(2013)]%
        {Rossow2013}
\bibfield{author}{\bibinfo{person}{Christian Rossow}, \bibinfo{person}{Dennis
  Andriesse}, \bibinfo{person}{Tillmann Werner}, \bibinfo{person}{Brett
  Stone-Gross}, \bibinfo{person}{Daniel Plohmann},
  \bibinfo{person}{Christian~J. Dietrich}, {and} \bibinfo{person}{Herbert
  Bos}.} \bibinfo{year}{2013}\natexlab{}.
\newblock \showarticletitle{SoK: P2PWNED - Modeling and Evaluating the
  Resilience of Peer-to-Peer Botnets}. In \bibinfo{booktitle}{\emph{Symposium
  on Security and Privacy}}. \bibinfo{publisher}{IEEE}.
\newblock
\urldef\tempurl%
\url{https://doi.org/10.1109/SP.2013.17}
\showDOI{\tempurl}


\bibitem[Saxe and Berlin(2017)]%
        {saxe_expose_2017}
\bibfield{author}{\bibinfo{person}{Joshua Saxe} {and}
  \bibinfo{person}{Konstantin Berlin}.} \bibinfo{year}{2017}\natexlab{}.
\newblock \showarticletitle{{eXpose}: {A Character-Level Convolutional Neural
  Network with Embeddings For Detecting Malicious {URLs}, File Paths and
  Registry Keys}}. \bibinfo{publisher}{arXiv:1702.08568}.
\newblock
\urldef\tempurl%
\url{https://doi.org/10.48550/ARXIV.1702.08568}
\showDOI{\tempurl}


\bibitem[Schiavoni et~al\mbox{.}(2014)]%
        {schiavoni_phoenix_2014}
\bibfield{author}{\bibinfo{person}{Stefano Schiavoni},
  \bibinfo{person}{Federico Maggi}, \bibinfo{person}{Lorenzo Cavallaro}, {and}
  \bibinfo{person}{Stefano Zanero}.} \bibinfo{year}{2014}\natexlab{}.
\newblock \showarticletitle{{Phoenix: {DGA}-Based Botnet Tracking and
  Intelligence}}. In \bibinfo{booktitle}{\emph{Detection of Intrusions and
  Malware, and Vulnerability Assessment}}. \bibinfo{publisher}{{Springer}}.
\newblock
\newblock
\shownote{\url{https://doi.org/10.1007/978-3-319-08509-8_11}}.


\bibitem[Sch{\"u}ppen et~al\mbox{.}(2018)]%
        {schuppen_fanci_2018}
\bibfield{author}{\bibinfo{person}{Samuel Sch{\"u}ppen},
  \bibinfo{person}{Dominik Teubert}, \bibinfo{person}{Patrick Herrmann}, {and}
  \bibinfo{person}{Ulrike Meyer}.} \bibinfo{year}{2018}\natexlab{}.
\newblock \showarticletitle{{FANCI : Feature-based Automated NXDomain
  Classification and Intelligence}}. In \bibinfo{booktitle}{\emph{{{USENIX
  Security Symposium}}}}. \bibinfo{publisher}{USENIX Association}.
\newblock
\newblock
\shownote{\url{https://www.usenix.org/conference/usenixsecurity18/presentation/schuppen}}.


\bibitem[Shi et~al\mbox{.}(2018)]%
        {shi_malicious_2018}
\bibfield{author}{\bibinfo{person}{Yong Shi}, \bibinfo{person}{Gong Chen},
  {and} \bibinfo{person}{Juntao Li}.} \bibinfo{year}{2018}\natexlab{}.
\newblock \showarticletitle{{Malicious Domain Name Detection Based on Extreme
  Machine Learning}}.
\newblock \bibinfo{journal}{\emph{Neural Processing Letters}}
  \bibinfo{volume}{48}, \bibinfo{number}{3} (\bibinfo{year}{2018}).
\newblock
\showISSN{1573-773X}
\newblock
\shownote{\url{https://doi.org/10.1007/s11063-017-9666-7}}.


\bibitem[Shu et~al\mbox{.}(2021)]%
        {Shu2022}
\bibfield{author}{\bibinfo{person}{Xiang Shu}, \bibinfo{person}{Chunjie Cao},
  \bibinfo{person}{Longjuan Wang}, {and} \bibinfo{person}{Fangjian Tao}.}
  \bibinfo{year}{2021}\natexlab{}.
\newblock \showarticletitle{{GWDGA:} An Effective Adversarial {DGA}}. In
  \bibinfo{booktitle}{\emph{Frontiers in Cyber Security}}.
  \bibinfo{publisher}{Springer}.
\newblock
\urldef\tempurl%
\url{https://doi.org/10.1007/978-981-19-0523-0\_3}
\showDOI{\tempurl}


\bibitem[Sidi et~al\mbox{.}(2020)]%
        {Sidi2019}
\bibfield{author}{\bibinfo{person}{Lior Sidi}, \bibinfo{person}{Asaf Nadler},
  {and} \bibinfo{person}{Asaf Shabtai}.} \bibinfo{year}{2020}\natexlab{}.
\newblock \showarticletitle{MaskDGA: An Evasion Attack Against DGA Classifiers
  and Adversarial Defenses}.
\newblock \bibinfo{journal}{\emph{IEEE Access}}  \bibinfo{volume}{8}
  (\bibinfo{year}{2020}).
\newblock
\urldef\tempurl%
\url{https://doi.org/10.1109/ACCESS.2020.3020964}
\showDOI{\tempurl}


\bibitem[Sivaguru et~al\mbox{.}(2018)]%
        {sivaguru_evaluation_2018}
\bibfield{author}{\bibinfo{person}{Raaghavi Sivaguru}, \bibinfo{person}{Chhaya
  Choudhary}, \bibinfo{person}{Bin Yu}, \bibinfo{person}{Vadym Tymchenko},
  \bibinfo{person}{Anderson Nascimento}, {and} \bibinfo{person}{Martine~De
  Cock}.} \bibinfo{year}{2018}\natexlab{}.
\newblock \showarticletitle{{An Evaluation of DGA Classifiers}}. In
  \bibinfo{booktitle}{\emph{International Conference on Big Data}}.
  \bibinfo{publisher}{IEEE}.
\newblock
\newblock
\shownote{\url{https://doi.org/10.1109/BigData.2018.8621875}}.


\bibitem[Spooren et~al\mbox{.}(2019)]%
        {Spooren2019}
\bibfield{author}{\bibinfo{person}{Jan Spooren}, \bibinfo{person}{Davy
  Preuveneers}, \bibinfo{person}{Lieven Desmet}, \bibinfo{person}{Peter
  Janssen}, {and} \bibinfo{person}{Wouter Joosen}.}
  \bibinfo{year}{2019}\natexlab{}.
\newblock \showarticletitle{Detection of Algorithmically Generated Domain Names
  Used by Botnets: A Dual Arms Race}. In \bibinfo{booktitle}{\emph{Symposium on
  Applied Computing}}. \bibinfo{publisher}{ACM}.
\newblock
\showISBNx{9781450359337}
\urldef\tempurl%
\url{https://doi.org/10.1145/3297280.3297467}
\showDOI{\tempurl}


\bibitem[Szegedy et~al\mbox{.}(2014)]%
        {Szegedy2013}
\bibfield{author}{\bibinfo{person}{Christian Szegedy},
  \bibinfo{person}{Wojciech Zaremba}, \bibinfo{person}{Ilya Sutskever},
  \bibinfo{person}{Joan Bruna}, \bibinfo{person}{Dumitru Erhan},
  \bibinfo{person}{Ian~J. Goodfellow}, {and} \bibinfo{person}{Rob Fergus}.}
  \bibinfo{year}{2014}\natexlab{}.
\newblock \showarticletitle{Intriguing properties of neural networks}. In
  \bibinfo{booktitle}{\emph{International Conference on Learning
  Representations}}. \bibinfo{publisher}{OpenReview.net}.
\newblock
\urldef\tempurl%
\url{https://openreview.net/forum?id=kklr_MTHMRQjG}
\showURL{%
\tempurl}


\bibitem[{The Domain Name Industry Brief}(2023)]%
        {dnibQ3-2023}
\bibfield{author}{\bibinfo{person}{{The Domain Name Industry Brief}}.}
  \bibinfo{year}{2023}\natexlab{}.
\newblock \bibinfo{title}{The Domain Name Industry Brief Q3 2023}.
\newblock
\newblock
\urldef\tempurl%
\url{https://dnib.com/articles/the-domain-name-industry-brief-q3-2023}
\showURL{%
\tempurl}
\newblock
\shownote{online, accessed 2023-12-02}.


\bibitem[Tong et~al\mbox{.}(2020)]%
        {tong_far_2020}
\bibfield{author}{\bibinfo{person}{Mingkai Tong}, \bibinfo{person}{Guo Li},
  \bibinfo{person}{Runzi Zhang}, \bibinfo{person}{Jianxin Xue},
  \bibinfo{person}{Wenmao Liu}, {and} \bibinfo{person}{Jiahai Yang}.}
  \bibinfo{year}{2020}\natexlab{}.
\newblock \showarticletitle{{Far from Classification Algorithm: Dive into the
  Preprocessing Stage in DGA Detection}}. In
  \bibinfo{booktitle}{\emph{International Conference on Trust, Security and
  Privacy in Computing and Communications}}. \bibinfo{publisher}{IEEE}.
\newblock
\newblock
\shownote{\url{https://doi.org/10.1109/TrustCom50675.2020.00070}}.


\bibitem[Tramer et~al\mbox{.}(2020)]%
        {Tramer2020}
\bibfield{author}{\bibinfo{person}{Florian Tramer}, \bibinfo{person}{Nicholas
  Carlini}, \bibinfo{person}{Wieland Brendel}, {and}
  \bibinfo{person}{Aleksander Madry}.} \bibinfo{year}{2020}\natexlab{}.
\newblock \showarticletitle{On Adaptive Attacks to Adversarial Example
  Defenses}. In \bibinfo{booktitle}{\emph{Advances in Neural Information
  Processing Systems}}. \bibinfo{publisher}{Curran Associates, Inc.}
\newblock
\urldef\tempurl%
\url{https://proceedings.neurips.cc/paper/2020/hash/11f38f8ecd71867b42433548d1078e38-Abstract.html}
\showURL{%
\tempurl}


\bibitem[Tran et~al\mbox{.}(2018)]%
        {tran_lstm_2018}
\bibfield{author}{\bibinfo{person}{Duc Tran}, \bibinfo{person}{Hieu Mac},
  \bibinfo{person}{Van Tong}, \bibinfo{person}{Hai~Anh Tran}, {and}
  \bibinfo{person}{Linh~Giang Nguyen}.} \bibinfo{year}{2018}\natexlab{}.
\newblock \showarticletitle{{A LSTM based framework for handling multiclass
  imbalance in DGA botnet detection}}.
\newblock \bibinfo{journal}{\emph{Neurocomputing}}  \bibinfo{volume}{275}
  (\bibinfo{year}{2018}).
\newblock
\urldef\tempurl%
\url{https://doi.org/10.1016/j.neucom.2017.11.018}
\showDOI{\tempurl}


\bibitem[Wang et~al\mbox{.}(2020)]%
        {Wang2020}
\bibfield{author}{\bibinfo{person}{Tianlu Wang}, \bibinfo{person}{Xuezhi Wang},
  \bibinfo{person}{Yao Qin}, \bibinfo{person}{Ben Packer},
  \bibinfo{person}{Kang Li}, \bibinfo{person}{Jilin Chen},
  \bibinfo{person}{Alex Beutel}, {and} \bibinfo{person}{Ed Chi}.}
  \bibinfo{year}{2020}\natexlab{}.
\newblock \showarticletitle{{CAT}-Gen: Improving Robustness in {NLP} Models via
  Controlled Adversarial Text Generation}. In
  \bibinfo{booktitle}{\emph{Empirical Methods in Natural Language Processing}}.
  \bibinfo{publisher}{Association for Computational Linguistics}.
\newblock
\urldef\tempurl%
\url{https://doi.org/10.18653/v1/2020.emnlp-main.417}
\showDOI{\tempurl}


\bibitem[Woodbridge et~al\mbox{.}(2016)]%
        {woodbridge_predicting_2016}
\bibfield{author}{\bibinfo{person}{Jonathan Woodbridge},
  \bibinfo{person}{Hyrum~S. Anderson}, \bibinfo{person}{Anjum Ahuja}, {and}
  \bibinfo{person}{Daniel Grant}.} \bibinfo{year}{2016}\natexlab{}.
\newblock \showarticletitle{{Predicting Domain Generation Algorithms with Long
  Short-Term Memory Networks}}. \bibinfo{publisher}{arXiv:1611.00791}.
\newblock
\urldef\tempurl%
\url{https://doi.org/10.48550/ARXIV.1611.00791}
\showDOI{\tempurl}


\bibitem[Wu et~al\mbox{.}(2020)]%
        {Wu2020}
\bibfield{author}{\bibinfo{person}{Zuxuan Wu}, \bibinfo{person}{Ser{-}Nam Lim},
  \bibinfo{person}{Larry~S. Davis}, {and} \bibinfo{person}{Tom Goldstein}.}
  \bibinfo{year}{2020}\natexlab{}.
\newblock \showarticletitle{Making an Invisibility Cloak: Real World
  Adversarial Attacks on Object Detectors}. In
  \bibinfo{booktitle}{\emph{European Conference on Computer Vision}}.
  \bibinfo{publisher}{Springer}.
\newblock
\urldef\tempurl%
\url{https://doi.org/10.1007/978-3-030-58548-8\_1}
\showDOI{\tempurl}


\bibitem[Yadav and Reddy(2012)]%
        {yadav_winning_2012}
\bibfield{author}{\bibinfo{person}{Sandeep Yadav} {and}
  \bibinfo{person}{A.~L.~Narasimha Reddy}.} \bibinfo{year}{2012}\natexlab{}.
\newblock \showarticletitle{{Winning with DNS Failures: Strategies for Faster
  Botnet Detection}}. In \bibinfo{booktitle}{\emph{Security and Privacy in
  Communication Networks}}. \bibinfo{publisher}{Springer}.
\newblock
\newblock
\shownote{\url{https://doi.org/10.1007/978-3-642-31909-9_26}}.


\bibitem[Yang et~al\mbox{.}(2022)]%
        {Yang2022}
\bibfield{author}{\bibinfo{person}{Yijun Yang}, \bibinfo{person}{Ruiyuan Gao},
  \bibinfo{person}{Yu Li}, \bibinfo{person}{Qiuxia Lai}, {and}
  \bibinfo{person}{Qiang Xu}.} \bibinfo{year}{2022}\natexlab{}.
\newblock \showarticletitle{What You See is Not What the Network Infers:
  Detecting Adversarial Examples Based on Semantic Contradiction}. In
  \bibinfo{booktitle}{\emph{Network and Distributed System Security
  Symposium}}. \bibinfo{publisher}{Internet Society}.
\newblock
\urldef\tempurl%
\url{https://doi.org/10.14722/ndss.2022.24001}
\showDOI{\tempurl}


\bibitem[Yoachimik(2022)]%
        {Yoachimik2022}
\bibfield{author}{\bibinfo{person}{Omer Yoachimik}.}
  \bibinfo{year}{2022}\natexlab{}.
\newblock \bibinfo{title}{Mantis - the most powerful botnet to date}.
\newblock
\newblock
\urldef\tempurl%
\url{https://blog.cloudflare.com/mantis-botnet/}
\showURL{%
\tempurl}
\newblock
\shownote{online, accessed 2023-11-21}.


\bibitem[Yoo et~al\mbox{.}(2020)]%
        {Yoo2020}
\bibfield{author}{\bibinfo{person}{Jin~Yong Yoo}, \bibinfo{person}{John
  Morris}, \bibinfo{person}{Eli Lifland}, {and} \bibinfo{person}{Yanjun Qi}.}
  \bibinfo{year}{2020}\natexlab{}.
\newblock \showarticletitle{Searching for a Search Method: Benchmarking Search
  Algorithms for Generating {NLP} Adversarial Examples}. In
  \bibinfo{booktitle}{\emph{BlackboxNLP Workshop on Analyzing and Interpreting
  Neural Networks for NLP}}. \bibinfo{publisher}{Association for Computational
  Linguistics}.
\newblock
\urldef\tempurl%
\url{https://doi.org/10.18653/v1/2020.blackboxnlp-1.30}
\showDOI{\tempurl}


\bibitem[Yoo and Qi(2021)]%
        {Yoo2021}
\bibfield{author}{\bibinfo{person}{Jin~Yong Yoo} {and} \bibinfo{person}{Yanjun
  Qi}.} \bibinfo{year}{2021}\natexlab{}.
\newblock \showarticletitle{Towards Improving Adversarial Training of {NLP}
  Models}. In \bibinfo{booktitle}{\emph{Findings of the Association for
  Computational Linguistics: EMNLP 2021}}. \bibinfo{publisher}{Association for
  Computational Linguistics}.
\newblock
\urldef\tempurl%
\url{https://doi.org/10.18653/v1/2021.findings-emnlp.81}
\showDOI{\tempurl}


\bibitem[Yu et~al\mbox{.}(2018)]%
        {yu_character_2018}
\bibfield{author}{\bibinfo{person}{Bin Yu}, \bibinfo{person}{Jie Pan},
  \bibinfo{person}{Jiaming Hu}, \bibinfo{person}{Anderson Nascimento}, {and}
  \bibinfo{person}{Martine De~Cock}.} \bibinfo{year}{2018}\natexlab{}.
\newblock \showarticletitle{{Character Level based Detection of DGA Domain
  Names}}. In \bibinfo{booktitle}{\emph{International Joint Conference on
  Neural Networks}}. \bibinfo{publisher}{{IEEE}}.
\newblock
\newblock
\shownote{\url{https://doi.org/10.1109/IJCNN.2018.8489147}}.


\bibitem[Yun et~al\mbox{.}(2020)]%
        {Yun2019}
\bibfield{author}{\bibinfo{person}{Xiaochun Yun}, \bibinfo{person}{Ji Huang},
  \bibinfo{person}{Yipeng Wang}, \bibinfo{person}{Tianning Zang},
  \bibinfo{person}{Yuan Zhou}, {and} \bibinfo{person}{Yongzheng Zhang}.}
  \bibinfo{year}{2020}\natexlab{}.
\newblock \showarticletitle{Khaos: An Adversarial Neural Network DGA With High
  Anti-Detection Ability}.
\newblock \bibinfo{journal}{\emph{Transactions on Information Forensics and
  Security}}  \bibinfo{volume}{15} (\bibinfo{year}{2020}).
\newblock
\urldef\tempurl%
\url{https://doi.org/10.1109/TIFS.2019.2960647}
\showDOI{\tempurl}


\bibitem[Zang et~al\mbox{.}(2020)]%
        {Zang2019}
\bibfield{author}{\bibinfo{person}{Yuan Zang}, \bibinfo{person}{Fanchao Qi},
  \bibinfo{person}{Chenghao Yang}, \bibinfo{person}{Zhiyuan Liu},
  \bibinfo{person}{Meng Zhang}, \bibinfo{person}{Qun Liu}, {and}
  \bibinfo{person}{Maosong Sun}.} \bibinfo{year}{2020}\natexlab{}.
\newblock \showarticletitle{Word-level Textual Adversarial Attacking as
  Combinatorial Optimization}. In \bibinfo{booktitle}{\emph{Annual Meeting of
  the Association for Computational Linguistics}}.
  \bibinfo{publisher}{Association for Computational Linguistics}.
\newblock
\urldef\tempurl%
\url{https://doi.org/10.18653/v1/2020.acl-main.540}
\showDOI{\tempurl}


\bibitem[Zhai et~al\mbox{.}(2022)]%
        {Zhai2022}
\bibfield{author}{\bibinfo{person}{You Zhai}, \bibinfo{person}{Jian Yang},
  \bibinfo{person}{Zixiang Wang}, \bibinfo{person}{Longtao He},
  \bibinfo{person}{Liqun Yang}, {and} \bibinfo{person}{Zhoujun Li}.}
  \bibinfo{year}{2022}\natexlab{}.
\newblock \showarticletitle{Cdga: A GAN-based Controllable Domain Generation
  Algorithm}. In \bibinfo{booktitle}{\emph{International Conference on Trust,
  Security and Privacy in Computing and Communications}}.
  \bibinfo{publisher}{IEEE}.
\newblock
\urldef\tempurl%
\url{https://doi.org/10.1109/TrustCom56396.2022.00056}
\showDOI{\tempurl}


\bibitem[Zheng et~al\mbox{.}(2021)]%
        {Zheng2021}
\bibfield{author}{\bibinfo{person}{Yu Zheng}, \bibinfo{person}{Chao Yang},
  \bibinfo{person}{Yanzhou Yang}, \bibinfo{person}{Qixian Ren},
  \bibinfo{person}{Yue Li}, {and} \bibinfo{person}{Jianfeng Ma}.}
  \bibinfo{year}{2021}\natexlab{}.
\newblock \showarticletitle{ShadowDGA: Toward Evading DGA Detectors with GANs}.
  In \bibinfo{booktitle}{\emph{International Conference on Computer
  Communications and Networks}}. \bibinfo{publisher}{IEEE}.
\newblock
\urldef\tempurl%
\url{https://doi.org/10.1109/ICCCN52240.2021.9522282}
\showDOI{\tempurl}


\bibitem[Zhu et~al\mbox{.}(2020)]%
        {Zhu2019}
\bibfield{author}{\bibinfo{person}{Chen Zhu}, \bibinfo{person}{Yu Cheng},
  \bibinfo{person}{Zhe Gan}, \bibinfo{person}{Siqi Sun}, \bibinfo{person}{Tom
  Goldstein}, {and} \bibinfo{person}{Jingjing Liu}.}
  \bibinfo{year}{2020}\natexlab{}.
\newblock \showarticletitle{FreeLB: Enhanced Adversarial Training for Natural
  Language Understanding}. In \bibinfo{booktitle}{\emph{International
  Conference on Learning Representations}}.
  \bibinfo{publisher}{OpenReview.net}.
\newblock
\urldef\tempurl%
\url{https://openreview.net/forum?id=BygzbyHFvB}
\showURL{%
\tempurl}


\bibitem[Zimmermann et~al\mbox{.}(2022)]%
        {Zimmermann2022}
\bibfield{author}{\bibinfo{person}{Roland~S. Zimmermann},
  \bibinfo{person}{Wieland Brendel}, \bibinfo{person}{Florian Tramer}, {and}
  \bibinfo{person}{Nicholas Carlini}.} \bibinfo{year}{2022}\natexlab{}.
\newblock \showarticletitle{Increasing Confidence in Adversarial Robustness
  Evaluations}. In \bibinfo{booktitle}{\emph{Advances in Neural Information
  Processing Systems}}. \bibinfo{publisher}{Curran Associates, Inc.}
\newblock
\urldef\tempurl%
\url{https://proceedings.neurips.cc/paper_files/paper/2022/hash/5545d9bcefb7d03d5ad39a905d14fbe3-Abstract-Conference.html}
\showURL{%
\tempurl}


\end{thebibliography}

\appendix
\section{Additional Information on Attacks}
\label{sec:appendix}

\subsection{Embedding-Space Attacks}
\label{sec:appendix_embedding_space_attacks} 
The research community has developed many adversarial attacks.
Some of the most influential attacks include:
The \emph{Fast Gradient Sign Method (FGSM)}~\cite{Goodfellow2014} that is one of the first gradient-based adversarial attacks generating input perturbations in one gradient ascend step.
The \emph{Basic Iterative Method (BIM)}~\cite{Kurakin2018} is an iterative version of the FGSM attack. 
\emph{Projected Gradient Descent (PGD)}~\cite{Madry2017} is an advanced iterative attack that uses a projection function in combination with random restarts to generate AEs.
It is considered one of the most powerful first-order attacks~\cite{Akhtar2021}.
\emph{Carlini and Wagner (C\&W)}~\cite{Carlini2017a} formulated the problem of generating AEs as an optimization problem and use the Adam~\cite{Kingma2017} optimizer to generate AEs with minimal perturbations.
Croce and Hein~\cite{Croce2020} identified hyperparameter selection as a major challenge of current adversarial attacks and proposed \emph{AutoAttack} as a parameter-free state-of-the-art ensemble attack to combat this issue.

For our robustness evaluation, we choose PGD both in its $L_2$ and $L_\infty$ variants, C\&W in its $L_2$ variant, and AutoAttack in its $L_2$ and $L_\infty$ variants, which we adapt to binary classifiers (BAT).
We explicitly choose not to evaluate against weak methods such as the FGSM as other attacks we include were demonstrated to be more powerful~\cite{Madry2017}, and we see no reason why an attacker would be able to use FGSM, but not PGD.
We also leave out attacks similar to the ones we choose, such as the BIM (which works similarly to PGD).

Many of our attack's (PGD $L_2$, PGD $L_\infty$, BAT $L_2$, BAT $L_\infty$) primary hyperparameter is the perturbation budget $\varepsilon$.
We evaluate the attacks for different values of $\varepsilon$ depending on the norm of the attack.
As our threat model allows for unbounded perturbations, we additionally perform our attacks in an unbounded scenario: $\varepsilon_\infty = 1$, and $\varepsilon_2 = \sqrt{63 * 128}$ (an embedded e2LD is a sequence of 128-dimensional vectors of length 63).
The C\&W attack does not require a perturbation budget.
It instead relies on the confidence hyperparameter $\kappa$.
We evaluate our attacks for ten different values of the described hyperparameters chosen on an approximately logarithmic scale. Table~\ref{tab:embedding_space_hyperparameters} shows the chosen hyperparameters.
\rowcolors{2}{white}{white}
\begin{table}
	\caption{Attack hyperparameters of embedding-space attacks.}
	\resizebox{\columnwidth}{!}{
		\begin{tabular}{c|c|c|c|c|c|c|c|c|c|c}
			$\varepsilon_\infty$ & 0.01 & 0.02 & 0.03 & 0.05 & 0.08 & 0.15 & 0.25 & 0.5 & 0.7 & 1 \\
			\hline
			$\varepsilon_2$ & 0.5 & 0.9 & 1.6 & 2.8 & 5 & 9 & 16 & 32 & 50 & $\sqrt{63 * 128}$ \\
			\hline
			$\kappa$ & 0 & 0.03 & 0.08 & 0.2 & 0.6 & 1.7 & 4.6 & 13 & 36 & 100 
		\end{tabular}
	}
	\label{tab:embedding_space_hyperparameters}
\end{table}
For the PGD and C\&W attacks, we  additionally set the number of iterations to $50$.
Our experiments show that our attacks are successful in this configuration, so we did not tune this parameter any further.

\subsection{Discretized Embedding-Space Attacks}
\label{sec:appendix_discretized_embedding_space_attacks} 
\begin{figure}
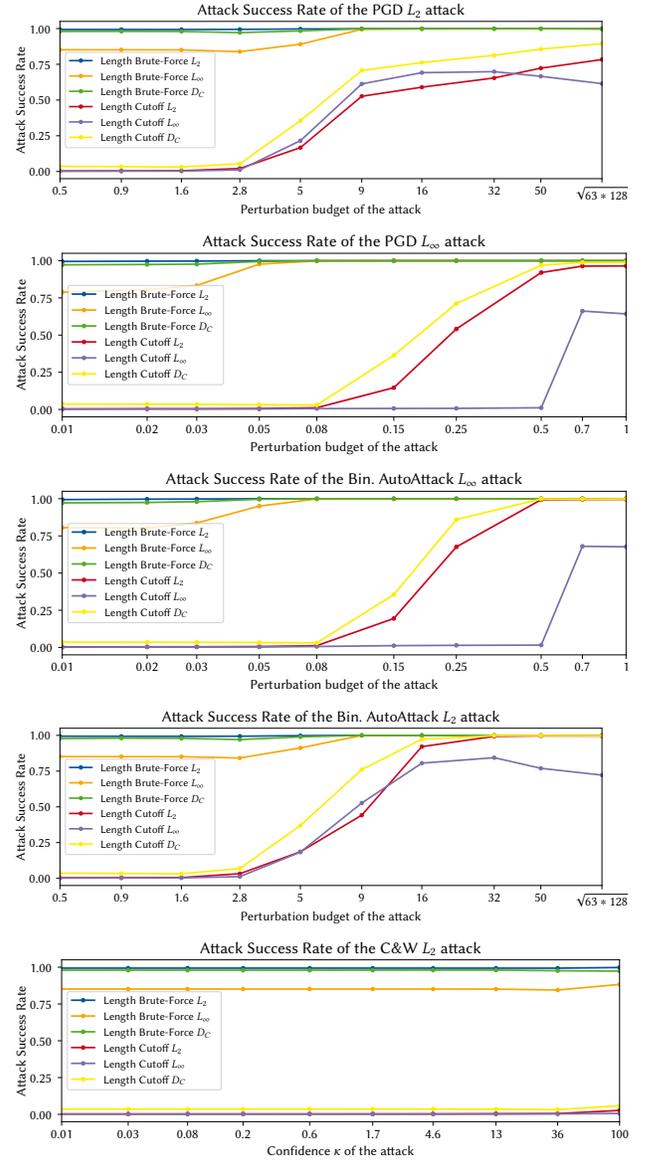

	\resizebox{\columnwidth}{!}{
		\centering
		\input{figures/fnr_under_pgd_l2_attack.pgf}
	}
	\resizebox{\columnwidth}{!}{
		\centering
		\input{figures/fnr_under_pgd_linf_attack.pgf}
	}
	\resizebox{\columnwidth}{!}{
		\centering
		\input{figures/fnr_under_bin_autoattack_l2.pgf}
	}
	\resizebox{\columnwidth}{!}{
		\centering
		\input{figures/fnr_under_bin_autoattack_linf.pgf}
	}
	\resizebox{\columnwidth}{!}{
		\centering
		\input{figures/fnr_under_cw_attack.pgf}
	}
	\caption{Success rates of the chosen attacks as a function of the attack strengths and discretization schemes.}
	\label{fig:Attacks}    
\end{figure}

In Fig.~\ref{fig:Attacks}, we present the success rates of the chosen attacks as a function of the attack strengths and discretization schemes.

Apart from the FNR, we also recorded additional metrics such as the percentage of unique domains, the distances to the original domains in the embedding space, the Levenshtein distances to the original domain, the model's confidence in the adversarial domains, and the percentage of useable adversarial domains.
We call an adversarial domain useable when it fools the classifier and has not been previously generated by the attack.
We present these metrics in Table~\ref{tab:EmbeddingSpaceRobustnessAttackKPIs}.

Across all our attacks, we observe that the Length Brute-Force (LBF) attacks are the strongest.
For all attacks but the C\&W $L_2$ attack, they generate more than $98.8\%$ useable adversarial domains in the strongest configuration of each attack. Additionally, the LBF attacks show very high FNRs even with low perturbation budgets.
However, the LBF $L_\infty$ attacks behave differently from the other attacks:
They only catch up with the other LBF attacks with higher perturbation budgets and do not catch up in the case of C\&W $L_2$ attacks.

The success of the Length-Cutoff (LCO) attacks is strongly coupled with the perturbation budget.
The attacks do not succeed unless the attack has been granted a sufficient perturbation budget to induce at least one change.
Additionally, we observe that LCO $L_\infty$ attacks only start being successful at very high perturbation budgets.
Furthermore, they are the only attacks whose success rate noticeably decreases at very high perturbation budgets, which is an indicator for a bad attack. 
As explained in Section~\ref{sec:desa}, LCO-based rounding is unsuitable in combination with the C\&W $L_2$ attack.

\rowcolors{3}{gray!20}{white}
\begin{table*}[t]
\footnotesize
  \caption{Results for the strongest configuration of each attack with every discretization scheme averaged over five folds.}
  \label{tab:EmbeddingSpaceRobustnessAttackKPIs}  
  \centering
  \begin{tabular}{ll|c|c|c|c|c|c|c}
    \textbf{Attack} & & \textbf{Useable} & \textbf{FNR} & \textbf{Conf.} & \textbf{\boldmath $L_2$ Dist.} & \textbf{\boldmath $L_\infty$ Dist.} & \textbf{Lev Dist.} & \textbf{Unique} \\
   \hline \Tstrut
    & LBF $L_2$ & 1.00000 & 1.00000 & 0.00000 & 14.37074 & 0.67887 & 51.18776 & 1.00000 \\
    \cellcolor{white} & LBF $L_\infty$ & 1.00000 & 1.00000 & 0.00001 & 11.62103 & 0.57568 & 45.52410 & 1.00000 \\
    & LBF $D_c$ & 1.00000 & 1.00000 & 0.00000 & 15.70944 & 0.70040 & 41.84694 & 1.00000 \\
    \cellcolor{white} & LCO $L_2$ & 0.96589 & 0.96604 & 0.05882 & 6.18322 & 0.55872 & 12.71709 & 0.99984 \\
    & LCO $L_\infty$ & 0.64374 & 0.64420 & 0.36603 & 6.32933 & 0.53696 & 14.47970 & 0.99910 \\
    \cellcolor{white} \multirow{-6}{*}{\textbf{PGD \boldmath $L_\infty$}} & LCO $D_c$ & 0.98882 & 0.99027 & 0.01782 & 6.80848 & 0.61945 & 12.98249 & 0.99855 \\
    \hline \Tstrut
     \cellcolor{gray!20} & LBF $L_2$ & 0.99729 & 0.99730 & 0.00338 & 11.14788 & 0.57472 & 42.26276 & 0.99999 \\ 
    & LBF $L_\infty$ & 0.99852 & 0.99852 & 0.00257 & 11.10966 & 0.55580 & 42.90393 & 1.00000 \\
    \cellcolor{gray!20}  & LBF $D_c$ & 0.99924 & 0.99925 & 0.00120 & 12.33196 & 0.66898 & 40.84781 & 0.99999 \\
    & LCO $L_2$ & 0.79592 & 0.79715 & 0.22498 & 5.20768 & 0.52011 & 10.10181 & 0.99675 \\
    \cellcolor{gray!20} & LCO $L_\infty$ & 0.62160 & 0.62171 & 0.38454 & 5.87969 & 0.52433 & 13.02559 & 0.99975 \\
    \multirow{-6}{*}{ \textbf{PGD \boldmath $L_2$}} & LCO $D_c$ & 0.89961 & 0.90111 & 0.11095 & 5.80743 & 0.57892 & 10.46043 & 0.99739 \\
   \hline \Tstrut
    & LBF $L_2$ & 0.99871 & 1.00000 & 0.00009 & 14.09738 & 0.67164 & 30.54480 & 0.99871 \\
    \cellcolor{white} & LBF $L_\infty$ & 0.99942 & 0.99955 & 0.00249 & 11.36193 & 0.55236 & 43.85597 & 0.99987 \\
    & LBF $D_c$ & 0.99283 & 1.00000 & 0.00027 & 13.85777 & 0.68355 & 26.57414 & 0.99283 \\
    \cellcolor{white} & LCO $L_2$ & 0.54026 & 0.99197 & 0.03641 & 6.34859 & 0.57082 & 12.12483 & 0.54795 \\
    & LCO $L_\infty$ & 0.54791 & 0.70460 & 0.32310 & 5.52853 & 0.48236 & 11.91279 & 0.75478 \\
    \cellcolor{white} \multirow{-6}{*}{\textbf{BAT \boldmath $L_\infty$}} & LCO $D_c$ & 0.50237 & 0.99728 & 0.02784 & 6.69565 & 0.59250 & 12.24533 & 0.50507 \\
   \hline \Tstrut
    \cellcolor{gray!20} & LBF $L_2$ & 0.99534 & 1.00000 & 0.00009 & 14.13603 & 0.67142 & 30.60022 & 0.99534 \\
    & LBF $L_\infty$ & 0.99942 & 0.99955 & 0.00249 & 11.47325 & 0.56196 & 44.29825 & 0.99987 \\
    \cellcolor{gray!20} & LBF $D_c$ & 0.98751 & 1.00000 & 0.00027 & 13.84007 & 0.68278 & 26.67205 & 0.98751 \\
    & LCO $L_2$ & 0.48477 & 0.99229 & 0.03789 & 6.31751 & 0.56617 & 12.12362 & 0.49194 \\
    \cellcolor{gray!20} & LCO $L_\infty$ & 0.56604 & 0.74558 & 0.28484 & 5.51901 & 0.48299 & 11.83996 & 0.74227 \\
    \multirow{-6}{*}{\textbf{BAT \boldmath $L_2$}} & LCO $D_c$ & 0.45012 & 0.99660 & 0.02954 & 6.64344 & 0.58819 & 12.24147 & 0.45329 \\
   \hline \Tstrut
    & LBF $L_2$ & 0.99679 & 0.99816 & 0.00501 & 9.51276 & 0.42862 & 36.39871 & 0.99863 \\
    \cellcolor{white} & LBF $L_\infty$ & 0.89082 & 0.89490 & 0.11117 & 8.71352 & 0.38773 & 29.67296 & 0.99143 \\
    & LBF $D_c$ & 0.97227 & 0.97592 & 0.03610 & 6.88070 & 0.51664 & 15.85198 & 0.99507 \\
    \cellcolor{white} & LCO $L_2$ & 0.12309 & 0.12325 & 0.82493 & 0.40586 & 0.05583 & 0.57539 & 0.99389 \\
    & LCO $L_\infty$ & 0.10436 & 0.10440 & 0.84366 & 0.07798 & 0.01540 & 0.05401 & 0.99849 \\
    \cellcolor{white} \multirow{-6}{*}{\textbf{CW \boldmath $L_2$}} & LCO $D_c$ & 0.15311 & 0.15337 & 0.79817 & 0.46609 & 0.06606 & 0.63116 & 0.99358 \\
    \end{tabular}  
\end{table*}  

\subsection{Discrete Attacks}
\label{sec:appendix_discrete_attacks}
Table~\ref{tbl:NLPRoustnessSummary} shows important KPIs of the tested discrete attacks.
\rowcolors{2}{gray!20}{white}
\begin{table}[t!]
	\caption{Results for the NLP attacks averaged over five folds.}
	\label{tbl:NLPRoustnessSummary} 
	\resizebox{\columnwidth}{!}{
		\begin{tabular}{l|c|c|c|c}
			\textbf{Attack} & \textbf{FNR} & \textbf{Unique} & \textbf{\boldmath $L_2$ dist.} & \textbf{\boldmath $L_\infty$ dist.} \\
			\hline \Tstrut
			HotFlip ($n = 1$) & 0.94043 & 0.99815 & 2.70735 & 0.56540 \\
			HotFlip ($n = 2$) & 0.99477 & 0.99686 & 3.89835 & 0.62637 \\
			HotFlip ($n = 3$) & 0.99719 & 0.99454 & 4.78771 & 0.65253 \\
			HotFlip ($n = 4$) & 0.99776 & 0.98942 & 5.51970 & 0.66712 \\
			HotFlip ($n = 5$) & 0.99800 & 0.97560 & 6.14546 & 0.67710 \\
			MaskDGA-WB & 0.99533 & 0.99597 & 6.45783 & 0.66530 \\
			\hline \Tstrut
			DeceptionDGA-BB & 0.76599 & 0.99910 & - & - \\
			DeepDGA-BB & 0.33968 & 1.00000 & - & - \\
			KhaosDGA-BB & 0.80410 & 1.00000 & - & - \\
			MaskDGA-BB & 0.77981 & 0.98240 & - & - \\
		\end{tabular}
	} 
\end{table}
Besides the FNR and the percentage of unique domains, we also list the mean $L_2$ and $L_\infty$ distance between the input domain and its adversarial domain in the embedding space.
We do not list the Levenshtein distance as it either matches the number of flips for HotFlip or half of the average domain length for MaskDGA-WB.

The table also shows the KPIs for the pre-computed black-box domains of DeceptionDGA, DeepDGA, KhaosDGA, and MaskDGA.
They are not particularly effective, only reaching a FNR of at most $80.4 \%$.
This is still impressive, considering that these domains were generated to fool different models trained on different data.

\rowcolors{2}{gray!20}{white}
\begin{table}[H]
	\caption{TPRs of the unhardened and hardened classifier on unknown DGAs.}
	\label{tbl:NovelDGAResults}
	\small
	\begin{tabular}{l|r|r}
		\textbf{DGA} & \textbf{TPR Baseline} & \textbf{TPR Hardened} \\
		\hline \Tstrut
		wd & 0.99961 & \textbf{1.00000} \\
		alien & \textbf{1.00000} & \textbf{1.00000} \\
		ccleaner & 0.91948 & \textbf{1.00000} \\
		tinynuke & 0.99920 & \textbf{1.00000} \\
		ares & 0.97200 & \textbf{1.00000} \\
		monerominer & 0.81246 & \textbf{0.99983} \\
		orchardgenesis & 0.97292 & \textbf{0.99931} \\
		darkwatchman & 0.97285 & \textbf{0.99877} \\
		necro & \textbf{0.99932} & 0.99873 \\
		orchard & 0.97293 & \textbf{0.99821} \\
		enviserv & 0.33360 & \textbf{0.99760} \\
		zloader & 0.98962 & \textbf{0.99349} \\
		flubot & 0.98732 & \textbf{0.99322} \\
		pseudomanuscrypt & 0.94560 & \textbf{0.96250} \\
		phorpiex & 0.93821 & \textbf{0.95911} \\
		dmsniff & 0.89905 & \textbf{0.94190} \\
		mydoom & 0.89798 & \textbf{0.91627} \\
		sharkbot & 0.83534 & \textbf{0.85257} \\
		kingminer & 0.76008 & \textbf{0.84044} \\
		bazarloader & 0.77629 & \textbf{0.83079} \\
		m0yv & \textbf{0.81351} & 0.81216 \\
		m0yvtdd & 0.80938 & \textbf{0.81212} \\
		g01 & 0.30192 & \textbf{0.34454} \\
		bigviktor & \textbf{0.32320} & 0.28760 \\
		nymaim2 & 0.11015 & \textbf{0.11263} \\
		qsnatch & \textbf{0.12244} & 0.08677 \\
		chaes & \textbf{0.00440} & 0.00330 \\
		\hline 
		\hline \Tstrut
		Average & 0.75810 & \textbf{0.80525} \\
	\end{tabular}
\end{table}

\begin{figure*}[p]  
	\centering
	\resizebox{!}{\textheight}{
		\input{figures/cross_fold_evaluations_plus.pgf}
	}
	\caption{FNRs of the adversarial training runs to investigate anomalies across all five folds (see. Section~\ref{sec:investigating_anomalies}).}
	\label{fig:ATCrossFoldPlus}
\end{figure*}

\end{document}